\documentstyle[aps,preprint]{revtex}
\newcommand{\Bbl}[1]{\mbox{\boldmath $#1$}}
\begin{document}
\draft
\title{Two dimensional electron transport in disordered \\
and ordered distributions of magnetic flux vortices}
\author{Mads Nielsen and Per Hedeg\aa{}rd}
\address{Niels Bohr Institute, \O{}rsted Laboratory, Universitetsparken 5,
2100 Copenhagen \O, Denmark}
\date{\today}
\maketitle
\begin{abstract}
We have considered the conductivity properties of a two dimensional electron
gas (2DEG) in two different kinds of inhomogeneous magnetic fields, i.e.\
a disordered distribution of magnetic flux vortices, and a periodic array of
magnetic flux vortices. The work falls in two parts. In the first part
we show how the phase shifts for an electron
scattering on an isolated vortex, can be calculated analytically, and related
to the transport properties through the differential cross section. In
the second part we present numerical results for the Hall conductivity
of the 2DEG in a periodic array of flux vortices found by
exact diagonalization. We find characteristic spikes in the Hall
conductance, when it is plotted against the filling fraction.
It is argued that the spikes can
be interpreted in terms of ``topological charge'' piling up across local and
global gaps in the energy spectrum.
\end{abstract}

\section{Introduction}
Over the last decade the two dimensional electron gas (2DEG) have been
exposed to a wide range of physical experiments, in which the electrons have
been perturbed by different configurations of electrostatic potentials, with
or without a homogeneous perpendicular magnetic field. These experiments
have shown new kinds of oscillations in the magnetoconductivity, with a
periodicity not given by the geometry of the Fermi surface,
as is the case with the Shubnikov-de Haas oscillations,
but given by the interaction of the two
length scales given respectively by the magnetic length, and by the
spatial structure of the potential, e.g.\ the Weiss
oscillations~\cite{WeissOsc}.
More recently, there have been increasing interest in systems
where the 2DEG is exposed to an {\em inhomogeneous}
perpendicular magnetic field. In such systems the inhomogeneities in the
magnetic field acts as perturbations of the 2DEG, relative to the homogeneous
magnetic field, where the band structure consists of the completely flat
Landau bands. The inhomogeneous magnetic field appears in the Hamiltonian
in the form of a non trivial vector potential. In the case of a periodic
variation in the magnetic field, it is possible to construct a periodic
vector potential, if and only if the flux through the unit cell of the field
is equal to a rational number, when measured in units of the flux quantum
$\phi_0=h/e$. Under these special circumstances the Hamiltonian is periodic,
and Bloch states can be used as a basis for the calculation of responce
properties of the electron gas.

In this paper we have considered a special class of spatially varying
magnetic fields which consists of flux vortices, that are either distributed
at random or placed in a regular lattice structure.
A system consisting of a 2DEG penetrated by a random distribution
of magnetic flux vortices, have been
experimentally realized by Geim et al.~\cite{Geim92,Geim93}.
They made a sandwich construction of
a GaAs/GaAlAs sample with a 2DEG at the interface, and a type~II
superconducting lead film, electrically disconnected from the 2DEG.
When the system was placed in an
external magnetic field, and cooled below the transition temperature of the
film, the magnetic field penetrated the film, and thereby the 2DEG, in the
form of Abrikosov vortices. When the external magnetic field is weak,
below 100G, the vortices will be well separated, and the 2DEG therefore sees
a very inhomogeneous magnetic field. In the experiments conducted by Geim
et al.\ the flux pinning in the film was strong, resulting in a disordered
distribution of flux vortices. This is the physical situation which we
investigate in Sec.~\ref{sec:single} below. In very clean films of type~II
superconducting material, the flux vortices will order in a periodic array,
i.e.\ an Abrikosov lattice, and thereby create a periodic magnetic field
at the 2DEG. This is the situation which we analyse in Sec.~\ref{sec:array}.

Several authors have investigated the transport properties of 2DEG's in
different kinds of inhomogeneous magnetic fields. Peeters and
Vasilopoulos~\cite{Peeters} have made a theoretical study of
the magnetoconductivity in a 2DEG in the presence of a magnetic field,
which was modulated weakly and periodically along one direction. They found
large oscillations in the longitudinal resistivity as a function of the
applied magnetic field strength. These oscillations are due to the
interference between the two length scales given respectively by the period
of the lateral variation of the magnetic field, and by the magnetic length
corresponding to the average background field. The oscillations are
reminicent of the Weiss oscillations, but have a higher amplitude and a
shifted phase, relative to the magnetoresistance oscillations induced by the
periodic electrostatic potential.

The problem of how the transport properties of the 2DEG is modified by the
presence of a random distribution of flux vortices, have been treated earlier
by A. V. Khaetskii~\cite{Khaetskii91}, and also by
Brey and Fertig~\cite{Brey}.
The approach used by these authors are basically similar to the one we have
presented in Sec.~\ref{sec:single}, i.e.\ based on the Boltzmann transport
equation. The main difference being that while Khaetskii have treated the
scattering in certain limiting cases, including the semiclassical, and Brey
and Fertig have calculated the scattering cross section numerically, we have
found an analytic expression for the scattering cross section for electrons
scattering on an idealized vortex. This has enabled us to study the
scattering in more detail, and to observe scattering resonances.

In Sec.~\ref{sec:array} we will address the ``paradox'' of how the Hall effect
can disappear in the following situation: We imagine a 2DEG in a regular
2D-lattice of flux vortices, with the magnetic field from a single vortex
exponentially damped with an exponential length $\xi$, in units of the lattice
spacing. We take the total flux from a single vortex to be $\phi_0/2$, as is
the case when the vortices come from a superconductor. When $\xi\gg 1$, the
field is homogeneous and the Hall conductivity is $\sigma_H =p\frac{e^2}{h}$,
where $p$ is the number of filled bands. In the other limit i.e.\ when
$\xi\rightarrow 0$, the time reversal symmetry is restored, and the Hall
conductivity vanish. The fact that the system has time reversal symmetry
when the vortices are infinitely thin, can be seen by subtracting a Dirac
string carying one quantum of magnetic flux $\phi_0$, from each flux vortex.
The introduction of the Dirac strings can not change any physical quantities,
and the procedure therefore establishes that the system, with infinitely thin
vortices with flux $+\phi_0/2$, is equivalent with the system with reversed
flux $-\phi_0/2$, through each vortex. The paradoxical situation arises
because it is known from general arguments~\cite{TKNN82}, that the
contribution to the total Hall conductivity from a single filled
nondegenerate band is a topological invariant, and therefore cannot change
gradually. The situation is even more clear cut if we imagine a periodic
array of exponential flux vortices carying one flux quantum $\phi_0$ each.
Then the limit $\xi=\infty$ corresponds to a homogeneous magnetic field,
while the opposite limit $\xi=0$ coresponds to free particles.
Incidentially this scheme can be used to establish an interpolating path
between the multifractal structure known as Hofstadters
butterfly~\cite{Hofstadter} which is a plot of the allowed energy levels for
electrons on a lattice in a homogeneous magnetic field, and the corresponding
plot for lattice-electrons in no field,
which is completely smooth~\cite{mads}.

The plan of the paper is as follows.
In Sec.~\ref{sec:single} we concentrate on the theory of electrons scattering
on a single vortex, and the physical consequences for the resistivities.
First we review the classical scattering theory in Sec.~\ref{sec:class},
before we discuss thre quantum theory of scattering in Sec.~\ref{sec:quantum}.
The longitudinal and transverse resistivities are discussed in
Sec.~\ref{sec:conduc}, and resonance scattering is demonstrated
in Sec.~\ref{sec:reso}.
The case of a 2DEG in a periodic array of flux
vortices, is the subject of Sec.~\ref{sec:array}. In the first part,
Sec.~\ref{sec:intro}--\ref{sec:bandcross}, of this
section the general theory of electron motion in a periodic magnetic field
is reviewed, and in the second part, Sec.~\ref{sec:num}--\ref{sec:sup},
we present the results of the numerical calculations.

\section{Single Vortex Scattering}\label{sec:single}

In this section we will consider the consequences of the introduction of
magnetic flux vortices into a 2 dimensional electron gas, in the approximation
where each vortex is treated as an individual scattering center. The vortices
are assumed to be distributed at random, homogeneously over the sample. The
average separation between the vortices is assumed so large, that we can
neglect interference from multiple scattering events. In the experimental
situation the mean free path $l_f$ from impurity and phonon scattering, may be
very long compared to the average separation between the vortices, due to the
very clean samples and liquid Helium temperature. This means that multiple
scattering and interference may have important consequences. Nevertheless we
will stick to the simplifying picture of vortices as individual scatterers
in this section. As we shall see, the gross features observed in experiments
on this system, can be accounted for within this approximation.

\subsection{The Classical Cross Section}\label{sec:class}

We will start by calculating the differential cross section for an electron
scattering on a flux vortex within the framework of classical mechanics. This
will provide a reference frame, and allow us to speak unambiguously about the
classical limit. In the calculations we shall use an ideal vortex, which has
a circular cross section with constant magnetic field inside, and zero
magnetic field outside
\begin{equation}\label{eq:bfield}
B(\bbox{r})=\left\{ \begin{array}{ccc}
B_0=\displaystyle{\frac{\phi}{\pi R_v^2}} & \mbox{for} &  r<R_v \\
0                          & \mbox{for} &  r>R_v .
\end{array}\right.
\end{equation}
Here $R_v$ is the radius, and $\phi$ is the total flux carried by the vortex.
The classical orbit is found as the
solution to Newton's equation of motion with the force given by the Lorentz
expression $\bbox{F}=-e\bbox{v\times B}$. It consists, as is well known, of
straight line segments outside the vortex, and an arc of a circle inside, with
radius of curvature given by the cyclotron radius $l_c =v/\omega_c$, with
$v$ being the particle velocity, and $\omega_c =\frac{eB}{m}$ the cyclotron
frequency. Inside the vortex the orbit is an arc of a circle, and as it is
impossible to draw a circle that only cut the circumference of the vortex
once, it is a simple geometrical consequence that a particle obeying the laws
of classical mechanics, and which initially is outside the vortex,
can never become trapped inside the vortex.
It is clear that the classical scattering is controlled by the single
parameter $\gamma=l_c/R_v$, which is the ratio between the radius of the
cyclotron orbit, and the radius of the flux vortex. Our first objective is
therefore, for a given $\gamma$, to find the relation between the impact
parameter $b$, and the scattering angle $\theta$.
Fig.~\ref{fig:geo} shows the geometry of the scattering, and the definition
of the impact parameter $b$, and angles $\phi,\psi,\theta$.
Let us define the reduced impact parameter $\beta=b/R_v$, which is bounded to
the interval $-1<\beta <1$. By inspecting Fig.~\ref{fig:geo} it is observed
that the following relations hold
\begin{eqnarray}
  \beta  & = & \sin \phi  \\
  \tan \psi & = & \frac{\gamma+\beta}{\sqrt{1-\beta^2}} \\
  \gamma\sin\frac{\theta}{2} & = & \sin (\psi - \phi)
  {\rm sign}(\gamma+\beta),
\end{eqnarray}
where the sign of $\gamma$ is dictated by the direction of the magnetic field
inside the vortex. We take $\gamma$ to be positive. After a small amount of
arithmetic
$\phi$ and $\psi$ are eliminated, and we have
\begin{equation}\label{eq:scatt}
\sin\frac{\theta}{2} = {\rm sign}(\gamma+\beta)
\sqrt{\frac{1-\beta^2}{\gamma^2 +2\gamma\beta +1}}.
\end{equation}
This relation gives the scattering angle as a function of the impact
parameter.
In classical scattering the scattering angle is always uniquely determined,
once the impact parameter is given, in contrast to the inverse, i.e.\ the
impact parameter as a function of the deflection angle.

In an experiment one would measure the number of particles per
time $\displaystyle{\frac{dN(\theta)}{dt}}d\theta$
scattered to an interval $d\theta$ about
the angle $\theta$. This will of course depend on the incoming flux of
particles $j$, defined as the number of particles per time that cross a unit
length perpendicular to the flow. Therefore we write
\begin{equation}
\frac{dN(\theta)}{dt}=j\frac{d\sigma}{d\theta}.
\end{equation}
The differential cross section $\frac{d\sigma}{d\theta}$ gives the total
weight of impact parameters, which give scattering into the direction
$\theta$. If, for a given angle $\theta$, we label the different values of the
impact parameter $b_1,b_2,\dots b_p$, which result in scattering into
$\theta$, then we have
\begin{equation}\label{eq:absb}
\frac{d\sigma}{d\theta}=\sum_{i=1}^{p}\left|\frac{db_i}{d\theta}\right|.
\end{equation}
Equation~\ref{eq:scatt} has at most two solutions, which are easily
found to be
\begin{equation}
\beta_{\pm}(\theta)=-\gamma\sin^2 \theta/2\; \pm\;
       \cos\theta/2 \sqrt{1-\gamma^2\sin^2\theta/2}.
\end{equation}
The solutions have to obey the auxiliary conditions $|\beta|\le 1$, and
sign$(\gamma+\beta)=$sign$(\theta)$. Furthermore we have
\begin{equation}
\frac{d\beta_{\pm}}{d\theta}= -\gamma \cos\theta/2\sin\theta/2 \mp
\sin\theta/2\frac{1+\gamma^2\cos\theta}{2\sqrt{1-\gamma^2\sin^2\theta/2}},
\end{equation}
from which the differential cross-section can be calculated from
Eqn.~\ref{eq:absb}, still having the auxiliary conditions in mind. Examples of
cross-sections and trajectories are shown in Fig.~\ref{fig:ccs}.
The integrated cross section
\begin{equation}
\sigma_{\rm tot}=\int_{-\pi}^{\pi}d\theta \left[\frac{d\sigma}{d\theta}\right],
\end{equation}
is equal to the total weight of impact parameters, which hit the
vortex. It is equal to the diameter of the vortex $\sigma_{\rm tot}=2R_v$, as
is always the case in classical scattering. Let us imagine an electron at the
Fermi surface scattering off the vortex. Then we have $l_c = v_F/\omega_c
=mv_F/eB=\hbar k_F/eB$. If we furthermore take the flux of the vortex to be
a fraction $f$ of the flux quantum $\phi_0$, the flux density becomes
$B=(f\phi_0)/\pi R_v^2=2f\hbar/e R_v^2$. The dimenensionless cyclotron radius
is then given by $\gamma=l_c/R_v=k_F R_v/2f$, and this parameter we call
$\kappa/2f$.
In the quantum regime $\kappa=2\pi R_v/\lambda_F$, and $f$ are the natural
parameters to characterize the scattering. This identification of parameters
allow us to compare the predictions of classical and quantum theory below.

\subsection{Quantum Scattering}\label{sec:quantum}

In this section we shall consider the electron scattering off a magnetic
flux vortex within the framework of quantum scattering theory. We will
calculate the differential cross section, from which the longitudinal and
transverse conductivities can be found from the theory of
Sec.~\ref{sec:conduc}. The quantum nature of the electron
radically alters the picture of the scattering
process, when the wavelength of the electron is comparable to, or longer
than the diameter of the vortex. In the limit of very small electron
wavelength, the scattering can essentially by described by the laws of
geometrical optics, and thereby classical mechanics.

We will again take an idealized cylindrical vortex, with constant magnetic
field inside, and zero field outside, Eqn.~\ref{eq:bfield}. This vortex is
completely symmetric under any rotation about the center axis. This symmetry
can also be made a symmetry of the Hamiltonian, by chosing a proper gauge
when writing down the vector potential. In cylindrical coordinates
$\bbox{A}=\bbox{e}_r A_r + \bbox{e}_{\theta} A_{\theta}$, we have
\begin{equation}
B(\bbox{r})=\partial_r A_{\theta} -\frac{1}{r}\partial_{\theta}A_r
+\frac{A_{\theta}}{r}.
\end{equation}
When $B$ is invariant under rotation, this equation has the simple solution
\begin{equation}
A_r=0,\;\;\;\;
A_{\theta}(r)=\frac{1}{r}\int_0^r dr'\, r'B(r'),
\end{equation}
which in our case give the vector potential $\bbox{A}=\bbox{e}_{\theta}
A_{\theta}$
with
\begin{equation}
A_{\theta}(r)=\left\{\begin{array}{ccc}
\displaystyle{\frac{\phi r}{2\pi R_v^2}}    &   {\rm for} & r<R_v \\
\displaystyle{\frac{\phi}{2\pi r}}      &  {\rm for} & r>R_v.
\end{array}\right.
\end{equation}
The Hamiltonian is given by the expression
\begin{equation}
H=\frac{1}{2m}\left(\bbox{p} + e\bbox{A}\right)^2
=-\frac{\hbar^2}{2m}\left\{\partial^2_r + \frac{1}{r}\partial_r +
\left(\frac{1}{r}\partial_{\theta} +
\frac{ie}{\hbar}A_{\theta}\right)^2\right\}.
\end{equation}
Here we have taken the charge of the electron to be $-e$. The Hamiltonian is
rotationally invariant, and therefore commutes with the angular momentum
about the symmetry axis, $L_z$. Consequently $L_z$ and $H$ have common
eigenstates. The canonical momentum of a charge-$q$
particle in a magnetic field is given by the expression
$\bbox{p}=m\bbox{v}+q\bbox{A}=\frac{\hbar}{i}\bbox{\nabla}$, and the operator
for
the angular momentum about the $z$-axis is $L_z=[ \bbox{r\times p}]_z=
\frac{\hbar}{i}\partial_{\theta}$. The eigenstates of $L_z$ are $e^{il\theta}$,
and the
requirement that the wave function have no cut when the vector potential is
non-singular, reduces the possible
values of $l$ to the set of positive and negative integers. We can now
separate the variables of the common eigenstates of $L_z$ and $H$, and write
\begin{equation}
\phi_{kl}(r,\theta)=R_{kl}(r)e^{il\theta},
\end{equation}
where $k$ is an energy label $E=\frac{\hbar^2 k^2}{2m}$.
Let us introduce the flux quantum $\phi_0 =h/e$ and the dimensionless
fraction $f=\phi/\phi_0$. The differential equations for the radial part of
the wave function, takes a particularly simple form if we write it down in
dimensionless variables $\xi=r/R_v$ and $\kappa=k R_v=2\pi R_v/\lambda$. The
energy variable $\kappa$ measures the size of the vortex compared to the
electron wavelength. In terms of $\kappa$ and $f$ the classical limit will be
$\kappa,f\gg 1$. With these definitions, the equation for the radial
part of the wave function for $\xi<1$ is
\begin{equation}
R'' + \frac{1}{\xi}R' +\left(\kappa^2 -
   (\frac{l}{\xi} + f\xi)^2\right) R =0.
\end{equation}
And for $\xi>1$ we have
\begin{equation}
R'' +\frac{1}{\xi}R' +\left(\kappa^2 - \frac{(l+f)^2}{\xi^2}\right) R=0.
\end{equation}
Inside the vortex an analytical solution to the radial equation can be found
by the following procedure
essentially due to L. Page~\cite{Page30,Hist}.
First we make the substitutions $\rho=\sqrt{2f}
\xi$ and $w=\frac{\kappa^2}{2f}$ (we assume $f>0$), which results in the
equation
\begin{equation}\label{eq:Rrho}
R'' + \frac{1}{\rho}R' + \left(w - \left(\frac{l}{\rho}+\frac{\rho}{2}
\right)^2\right)R=0,
\end{equation}
for $0<\rho<\sqrt{2f}$. Next we write the radial function as
\begin{equation}
R_l(\rho)=\rho^{m} e^{-\rho^2/4}V_l(\rho),
\end{equation}
with $m=|l|$, and insert this into Eqn.~\ref{eq:Rrho}. Hereby we get an
equation for $V_l$
\begin{equation}
V''_l + \left(\frac{2m+1}{\rho} - \rho\right)V'_l + (w - l - m -1)V_l=0.
\end{equation}
This equation can be further simplified by making the substitution
$x=\rho^2/2$. Finally we have the equation
\begin{equation}\label{eq:kummer}
x V''_l + (m+1 - x)V'_l - \frac{1}{2}(m+l+1-w)V_l=0.
\end{equation}
This differential equation belongs to a class of equations known as Kummer's
equation. Kummer's equation is a member of an even bigger class of equations
of the form $\sum_{p=0}^{n}(a_p + b_p x)\frac{d^p y}{dx^p}=0$, which can all
be solved by Laplace's method \cite{Landau}.
Kummer's equation is solved by the confluent hypergeometric functions
$M$ and $U$ (in the notation of Abramowitz and Stegun \cite{AS}).
The complete solution to Eqn.~\ref{eq:kummer} can be written
\begin{equation}
V_l(x)=c_1 M(\frac{1}{2}(l+m+1-w),m+1,x) + c_2 U(\frac{1}{2}(l+m+1-w),m+1,x).
\end{equation}
It turn out that in order that $R_l(\xi)$ be regular as
$\xi\rightarrow 0$, we must take $c_2=0$, and we can therefore write down
the solution to the Schr\"{o}dinger equation inside the vortex
\begin{equation}
\phi_{\kappa l}(\xi,\theta)=C_1 \xi^{|l|}e^{-\frac{1}{2}f\xi^2}
M(\frac{1}{2}(l+|l|+1 - \frac{\kappa^2}{2f}),|l|+1,f\xi^2)e^{il\theta}.
\end{equation}
Here $C_1$ is a normalization constant which we will not need to evaluate.

Outside the vortex the
radial equation is just the differential equation for ordinary Bessel
functions of the first kind. We therefore immediately have for $\xi>1$
\begin{equation}
\phi_{\kappa l}(\xi,\theta)=
\left(A_l J_{l+f}(\kappa \xi) + B_l Y_{l+f}(\kappa \xi)\right)e^{il\theta}.
\end{equation}
The two constants $A_l$, $B_l$ are found from the requirement, that the
wavefunction has to be continuously differentialble at the boundary of the
vortex.
There is no need to normalize the wave functions $\phi_{\kappa l}$, as the
normalization constant will drop out of the final expression.

In quantum scattering theory one seeks a particular eigenstate of the
Hamiltonian which belongs to the continnuous part of the spectrum, and
which far away from the scattering center has a direction in which it
represents an incoming flow of particle current.
Let us consider an eigenstate coresponding to the energy
$E_k=\frac{\hbar^2 k^2}{2m}$
\begin{equation}
\psi_k=\sum_l b_l \phi_{kl}.
\end{equation}
We want $\psi_k$ to represent the scattering of particles which are incident
along the x-axis, as indicated in Fig.~\ref{fig:qgeo}.
The asymptotic boundary condition on $\psi_k$ is therefore that it far to
the left of the origin represents an uniform current of incoming particles.

The vector potential
gives a contribution to the current $\bbox{j}$, so that the plane wave in the
direction of the $x$-axis is altered from the field free form
$e^{ikr\cos\theta}$. The particle current density is given by
\begin{equation}
\bbox{j}=\frac{\hbar}{2mi}\left\{ \Psi^{\dag}
\left[ (\bbox{\nabla}+\frac{ie}{\hbar}\bbox{A})\Psi\right] -
\left[ (\bbox{\nabla}+\frac{ie}{\hbar}\bbox{A})\Psi\right]^{\dag}\Psi\right\},
\end{equation}
and it is straightforward to check that the correct form for a state with
uniform current in the direction of the $x$-axis is
$e^{ikr\cos\theta - if\theta}$. If the flux through the vortex is not an
integer
number of flux quanta, i.e.\ if the fraction $f$ is not integer, then the
factor $e^{-if\theta}$ is not single valued as it stands, and we have to
introduce a cut to make it so. We only enforce the asymptotic boundary
condition along the negative $x$-axis, so the cut can be placed anywhere
outside this region.
Let us for a moment introduce the principal angle
$[\theta ]$, defined by $[\theta ]=\theta$ when $c<\theta<c+2\pi$, and
otherwise given by
periodicity. Then the single valued factor $e^{if[\theta ]}$ has a cut along
the
half line $\theta=c$. We want to express the plane wave, as a sum over partial
waves, and therefore we consider the inner product with $e^{il\theta}$, $l$
integer
\begin{eqnarray}
\int_{2\pi}\frac{d\theta}{2\pi}e^{ikr\cos\theta -if[\theta]-il\theta}=
\int_c^{c+2\pi}\frac{d\theta}{2\pi}e^{ikr\cos\theta -i(f+l)\theta} & = &
\nonumber \\
e^{i\frac{\pi}{2}(l+f)}\int_{c+\frac{\pi}{2}}^{c+\frac{5\pi}{2}}
         \frac{d\theta}{2\pi}e^{ikr\sin\theta -i(f+l)\theta} &=&
         i^{(l+f)}\bbox{J}_{l+f}(kr),
\end{eqnarray}
where the last equality sign holds if $c=-\frac{3\pi}{2}$.
This means that the cut
is placed along the positive $y$-axis, as shown in Fig.~\ref{fig:qgeo}.
The function $\bbox{J}_{\nu}(z)$ is known as Anger's function~\cite{GR}, and
coincide with Bessel's $J_{\nu}(z)$ when $\nu$ is an integer. For $\nu$ not
an integer the Anger function has the nice property that is goes
asymptotically as the Bessel function for large arguments, i.e.
\begin{equation}
\bbox{J}_{\nu}(z)=J_{\nu}(z) + \frac{\sin\pi\nu}{\pi z}\left[
1 - \frac{\nu}{z} + O(|z|^{-2})\right],
\end{equation}
for $z\longmapsto\infty$.
Let us remark that if the cut is placed somewhere else the integral will not
give Anger's function, but asymptotically it will still go as some
combination of Bessel functions.

Let us now subtract the plane wave from $\psi_k$
\begin{equation}
\psi_k (r,\theta) - e^{ikr\cos\theta - if\theta}=
- i^{l+f}\bbox{J}_{l+f}(\kappa \xi) +
b_l A_l J_{l+f}(\kappa \xi) +
b_l B_l Y_{l+f}(\kappa \xi).
\end{equation}
In the asymptotic region the involved functions can be expanded to give
\begin{eqnarray}
\psi_k(r,\theta) - e^{ikr\cos\theta - if\theta}=
\frac{1}{\sqrt{2\pi \kappa \xi}}
& \left\{ \left[ b_l A_l-i^{l+f} -ib_l B_l\right]
  e^{i \kappa \xi - i(l+f)\frac{\pi}{2} - i\frac{\pi}{4}}\right.
& +  \nonumber \\
& \left.\left[ b_l A_l-i^{l+f} +ib_l B_l\right]
  e^{-i \kappa \xi + i(l+f)\frac{\pi}{2} + i\frac{\pi}{4}}
\right\}.&
\end{eqnarray}
This combination of terms are the sum of an incoming and an outgoing circular
wave. The coefficient multiplying the incoming wave must vanish, as all
the ingoing current should be represented by the plane wave. We therefore
have the condition
\begin{equation}
b_l A_l-i^{l+f} +ib_l B_l =0.
\label{eq:con}
\end{equation}
The radial differential equations are real and linear, and therefore
$A_l$ and $B_l$ are real numbers by construction. The phase shifts
$\delta_l$ are defined by the relation
\begin{equation}
\delta_l=\arctan \frac{B_l}{A_l}.
\end{equation}
We note that the phase shifts are independent of the arbitrary normalization
of the wave functions. If we write $A_l=C_l \cos\delta_l$ and
$B_l=C_l \sin\delta_l$, then Eqn.~\ref{eq:con} can be solved to give
$b_l C_l = i^{l+f}e^{-i\delta_l}$. The outgoing circular wave which
represents the scattered current, is given by the expression
\begin{equation}
F(\theta) \frac{e^{ikr}}{\sqrt{r}} = \frac{e^{ikr}}{\sqrt{r/R_v}}
\sum_l {\cal F}_l e^{il\theta}.
\end{equation}
Putting things together, we get the
following expression
\begin{equation}\label{eq:fl}
{\cal F}_l=-\sqrt{\frac{2}{\pi \kappa}}
e^{i\frac{\pi}{4}-i\delta_l}\sin\delta_l .
\end{equation}
It is seen that ${\cal F}_l$ is a function of $l$ only through the phase
shifts $\delta_l$, we can therefore write ${\cal F}_l={\cal F}[\delta_l]$.
{}From the above expression we can in principle calculate ${\cal F}(\theta)$,
and
thereby the differential cross section $\frac{d\sigma}{d\theta}=|F(\theta)|^2$.
But this is only
practically feasible if the sum over $l$ converges so fast that we can
approximate it with a finite sum. The way things are stated above,
${\cal F}_l$ does
not go to zero for $l\rightarrow -\infty$, but rather goes to a constant
value. The reason for this is that we have not singled out the Aharonov-Bohm
contribution to the scattering amplitude ${\cal F}(\theta)$, which is of a
singular
nature. We will now pass to the Aharonov-Bohm limit, that is the limit where
$R_v\rightarrow 0$, while the flux is kept constant. We will then be able
to express the effect of the finite radius of the vortex, as the difference
in the scattering amplitude from the Aharonov-Bohm result
\begin{equation}
{\cal F}(\theta)=\delta {\cal F}(\theta) + {\cal F}^{AB}(\theta).
\end{equation}

When treating potential scattering for potentials with finite range, the
scattering wave function is directly written as the sum of the incoming
plane wave and the outgoing circular wave. The same thing can not be done
here, as can be seen by the fact that such a sum must have a cut, while the
true wave function can not have any cuts.
This is a consequence of the long range of the vector potential which far
away from the scattering center falls off as $1/r$.
The term $F(\theta)e^{ikr}/\sqrt{r}$ must therefore be interpreted as the
largest term in an asymptotic expansion of the part of the wave function
carrying the outward particle current \cite{Bohm}. This is not different from
the procedure used to calculate cross sections for scattering on long range
scalar potentials, i.e.\ Coulomb.

\subsubsection{The Aharonov-Bohm Limit}

In the limit of vanishing radius of the flux vortex $R_v=0$, we are left
with only one dimensionfull variable $k$, and therefore it is impossible to
express the scattering amplitude in dimensionless form.
When $R_v\rightarrow 0$ the eigenstates of the Hamiltonian are
everywhere given by the expression
\begin{equation}
\phi_{k l}(r,\theta)=
\left(A_l J_{l+f}(kr) + B_l Y_{l+f}(kr)\right)e^{il\theta}.
\end{equation}
The physical demand that
we have to impose on the solutions $\phi_{kl}$ is that of boundedness at
the origin. Let us put $\nu=l+f$, the order of the Bessel functions. The
properties of $J_{\nu}(z)$ and $Y_{\nu}(z)$ for $z\rightarrow 0$, we infer
from the relations
\begin{eqnarray}
J_{\nu}(z) & = & (\frac{1}{2}z)^{\nu}\sum_{k=0}^{\infty}
\frac{(-\frac{1}{4}z^2)^k}{k!\Gamma(\nu+k+1)} \\
Y_{\nu}(z) & = & \frac{J_{\nu}(z)\cos\nu\pi -J_{-\nu}(z)}{\sin\nu\pi},
\end{eqnarray}
valid for all values of $\nu$ and $z$. For $\nu>0$ $J_{\nu}$ is regular, and
$Y_{\nu}$ is irregular, and we therefore have to take $A_l=1$, $B_l=0$ for
$l+f>0$. For $\nu<0$ the condition for regularity is seen to be
$B_l/A_l=-\tan\pi\nu$. In order to select one of the two solutions of
$\tan\delta_l^{AB}=-\tan\pi (l+f)=-\tan\pi f$, we impose the additional
condition that
the radial wave function shall be positive around $r=0$. Which translates
into $(-1)^l\sin\delta_l^{AB}\sin\pi f<0$.
All in all this means that the phase shifts in
the Aharonov-Bohm limit are
\begin{equation}
\delta_l^{AB}=\left\{\begin{array}{ccc}
  0 & {\rm for} & l + f >0 \\
  -\pi (l+f)    & {\rm for} & l+f<0 \end{array}\right.
\end{equation}
modulo $2\pi$. We note that it is only the fractional part of $f$ which
play a role in this limit, any integer part may be absorbed in a redefinition
of $l$. We also note that the phase shifts are independent of the electron
wavelength. If we take $0<f<1$, the calculation of the scattering amplitude
look like this
\begin{equation}
F^{AB}(\theta)=\sqrt{\frac{2}{\pi k}}e^{i\frac{\pi}{4}+i\pi f}\sin\pi f
\sum_{l=1}^{\infty}e^{-il\theta}
=\frac{1}{\sqrt{2\pi k}}e^{-i\frac{\pi}{4}+i\pi f-i\frac{\theta}{2}}
\frac{\sin\pi f}{\sin\frac{\theta}{2}}.
\end{equation}
The differential cross section for scattering on an infinitely thin vortex
carrying a magnetic flux $f\phi_0$ is therefore
\begin{equation}
\left[\frac{d\sigma}{d\theta}\right]_{AB}=
\frac{1}{2\pi k}\frac{\sin^2\pi f}{\sin^2 \theta/2}.
\end{equation}
A result first derived by Y.\ Aharonov and D.\ Bohm in 1959, by a slightly
different approach~\cite{Bohm}.
The Aharonov-Bohm cross section corresponding to $f=1/2$ is plotted in
Fig.~\ref{fig:AB} for reference.
The AB cross section is completely symmetric
under reflection $\theta\mapsto -\theta$, for all values of $f$, and can not
give
rise to a net transverse
force on the electrons. On the other hand in analogy with ordinary impurities,
it can give rise to a finite lifetime of the electron states, and thereby
give a contribution to the longitudinal resistance of the system. We note that
the cross section is periodic in the flux, with period equal to the flux
quantum, and that it completely vanish when the flux is equal to an
integer number of flux quanta. This fact, that the scattering amplitude
$F^{AB}(\theta)$ disappers when the stringlike vortex contains an integer
number
of flux quanta, has useful consequences.

The AB cross section is non integrable because of the singularity at $\theta
=0$,
and we can therefore not calculate $\sigma_{tot}$. This is due to the long
range nature of the interaction, i.e.\  the vector potential, which fall
of only as $1/r$. This situation is not different from scattering on an
ordinary scalar potential with long range, such as the Coulomb potential.

\subsubsection{Phase Shifts for Scattering on Vortex with Finite
Radius}\label{subsec:phase}

In this section we will briefly describe how the phase shifts for scattering
on a cylindrical vortex with finite radius are calculated. The equation from
which the phase shifts are derived, is simply the equation which results from
the demand that the logarithmic derivative of the radial wave function must
be continuous at the boundary of the vortex
\begin{equation}
\frac{1}{R_l^{<}}\left.\frac{d R_l^<}{d\xi}\right|_{\xi=1}=
\frac{1}{R_l^>}\left.\frac{d R_l^{>}}{d\xi}\right|_{\xi=1}.
\end{equation}
Inside the vortex we have
\begin{equation}
E_l\equiv\frac{1}{R_l^{<}}\left.\frac{d R_l^<}{d\xi}\right|_{\xi=1}=
|l| -f + 2f\frac{a_l}{b_l}\frac{M(a_l+1,b_l+1,f)}{M(a_l,b_l,f)},
\end{equation}
where we have defined the parameters
$a_l=\frac{1}{2}(l+|l|+1 - \frac{\kappa^2}{2f})$, and $b_l=|l|+1$. Outside the
vortex the logarithmic derivative reads
\begin{equation}
\frac{1}{R_l^>}\left.\frac{d R_l^{>}}{d\xi}\right|_{\xi=1}=
\frac{j_l + y_l \tan\delta_l}{J_{l+f}(\kappa) + Y_{l+f}(\kappa)\tan\delta_l},
\end{equation}
where we have introduced the abbreviations
$j_l=\kappa J_{l+f-1}(\kappa) - (l+f)J_{l+f}(\kappa)$ and
$y_l=\kappa Y_{l+f-1}(\kappa) - (l+f)Y_{l+f}(\kappa)$.
It is now  simple to solve for $\delta_l$
\begin{equation}\label{eq:tangens}
\tan\delta_l =\frac{j_l - E_l J_{l+f}(\kappa)}{E_l Y_{l+f}(\kappa) - y_l}.
\end{equation}
The $\tan\delta_l$'s are the bricks from which cross sections and transport
coefficients can be build up.
The presented curves and cross sections have all been calculated with phase
shifts found by this expression with the help of {\em Mathematica}, which have
implementations of all the involved special functions.

\subsubsection{Asymmetric Scattering on Vortex with Finite Radius}

For a vortex with finite radius, we want to express the scattering
amplitude ${\cal F}(\theta)$ exclusively in terms of the dimensionless
parameters $\kappa$ and $f$.
The procedure for calculating the scattering amplitude, for scattering on a
vortex with finite radius, is as follows. First the $\delta {\cal F}_l$'s
are found from
\begin{equation}
\delta {\cal F}_l = {\cal F}[\delta_{\kappa l}] -
               {\cal F}[\delta^{AB}_l],
\end{equation}
where $\delta_{\kappa l}$ is the phase shift calculated numerically by
the formulas given in the above section, and
${\cal F}[\cdot]$ is given by Eqn.~\ref{eq:fl}.
For small values of $\kappa$ the $\delta {\cal F}_l$'s vanishes rapidly when
$|l|$ increases, making it possible to approximate the sum
$\sum_l \delta {\cal F}_l e^{il\theta}$ well by a finite number of terms.
We have found
that for $\kappa<10$, of order 20 terms are needed at most. The scattering
amplitude is then simply given by
\begin{equation}
{\cal F}(\theta)=\sum_{l=l_{\rm min}}^{l_{\rm max}}\delta {\cal F}_l
e^{il\theta}
  + {\cal F}^{AB}(\theta),
\end{equation}
where the dimensionless AB cross section is given by
${\cal F}^{AB}(\theta)=\sum_l {\cal F}[\delta^{AB}_l]e^{il\theta}$. From
which we get the dimensionless cross section
\begin{equation}
\frac{d\varsigma}{d\theta}=\left|{\cal F}(\theta)\right|^2.
\end{equation}
Plots of differential cross sections calculated by this procedure is
shown in Fig.~\ref{fig:asymcross}. We note that the degree of asymmetry is
determined by the size of the parameter $\kappa=k R_v$. The classical limit
is approached when $\kappa\gg 1$. The quantum cross section, unlike the
classical one, gives a finite probability of scattering to both sides of the
vortex for all parameter values, but not necessarily in all directions.

\subsection{Conductivity of 2DEG in Vortex Field}\label{sec:conduc}

In this section we will make an estimate of the contribution to the
resistance of a 2DEG, from a random distribution of flux vortices. The
experimental situation we have in mind, is that of Geim et.\
al.~\cite{Geim92,Geim93}, who placed a thin film of lead on top of a
GaAs/GaAlAs heterostructure.
When the temperature is lowered below the critical temperature,
and the magnetic field is below $H_{c2}$, the magnetic field penetrates the
superconductor in the form of Abrikosov vortices. At the 2DEG the field will
be confined to regions of radius (or exponential length) $\lambda_s$, each
threaded by a magnetic flux $\phi_0/2$. The vortices are assumed to be
distributed randomly, with an average separation $d$ given by the strength of
the external $B$-field according to the relationship $d^2 B_{\rm ext}=
\phi_0/2$.

The Boltzmann equation, linearized in the external electric field, reads
\begin{equation}\label{eq:boltz}
-e\bbox{v}\cdot\bbox{E}\frac{\partial f^0}{\partial \epsilon}=
\int\frac{d^2 q}{(2\pi)^2}\left\{f_{p+q}w_{p+q\rightarrow p} -
            f_{p}w_{p\rightarrow p+q}\right\}
   -\frac{f_{p} - f^0}{\tau_{imp}}.
\end{equation}
Here the transition probabilities $w_{p\rightarrow p+q}$ are potentially
asymmetrical quantities, due to the time reversal breaking magnetic field in
the vortices. As argued by B. I. Sturman~\cite{Sturman},
the correct form of the collision integral, even in the absence of
detailed balance, is the one given in Eqn.~\ref{eq:boltz}.
The electron-vortex scattering will be
elastic. In order to solve the Boltzmann equation we Fourier transform, and
write
\begin{eqnarray}
f(k,\theta) & = & \sum_{n=-\infty}^{\infty}e^{-in\theta}f_n(k) \\
w(k,\theta) & = & \sum_{n=-\infty}^{\infty}e^{-in\theta}w_n(k),
\end{eqnarray}
where $\theta$ is the angle between $\bbox{k}$ and $\bbox{E}$.
The electron-vortex collision integral is diagonal when Fourier transformed,
and we get the following equation for the $n$'th component of the
distribution function
\begin{equation}
-evE\frac{\partial
f^0}{\partial\epsilon}\frac{1}{2}\{\delta_{n,1}+\delta_{n,-1}\} =
-n_v \{w_0 - w_n\}f_n  -\frac{f_n - f^0\delta_{n,0}}{\tau_{imp}}.
\end{equation}
The current is given by
\begin{equation}\label{eq:current}
\bbox{j}=-2e\int\frac{d^2 k}{(2\pi)^2}\bbox{v}f(k,\theta)
=-\frac{e^2\epsilon_F E}{\pi \hbar^2 n_v}
\left\{\begin{array}{c}
{\rm Re}[\displaystyle{\frac{1}{w_1 - w_0}}] \\
{\rm Im}[\displaystyle{\frac{1}{w_1 - w_0}}]
\end{array}\right\},
\end{equation}
from which the conductivities can be read off. The resistivities are found by
inverting the conductivity tensor.

\subsubsection{Longitudinal Resistivity}

The longitudinal resistivity, obtained by the above procedure, is
\begin{equation}
\rho_{xx}=\frac{m}{ne^2}\left(\frac{1}{\tau_v} + \frac{1}{\tau_{imp}}\right),
\end{equation}
where the transport scattering time for the electrons, due to scattering on
the vortices, is
\begin{equation}\label{eq:txx}
\frac{1}{\tau_v}=n_v v_F \int_{-\pi}^{\pi}d\theta (1-\cos\theta)|F(\theta)|^2.
\end{equation}
Let us introduce a dimensionless quantity $\zeta$, by writing the
contribution to the longitudinal resistivity, from the electron-vortex
scattering, as
\begin{equation}
\rho^v_{xx}=\frac{\hbar}{e^2}\frac{n_v}{n}\zeta.
\end{equation}
The dimensionless parameter $\zeta$, which is explicitly given by the
expression
\begin{equation}
\zeta = \kappa\int_{-\pi}^{\pi}d\theta
(1-\cos\theta)\frac{d\varsigma}{d\theta},
\end{equation}
is characterizing the efficiency of a single vortex, to scatter the electrons
from the front to the back of the Fermi circle. It is straightforward to
do the integral and obtain the following sum
\begin{equation}
\zeta=4\sum_{l=-\infty}^{\infty}
  \frac{t_l(t_l - t_{l+1})}{(1+t_l^2)(1+t_{l+1}^2)},
\end{equation}
where $t_l=\tan\delta_l$ is found from Eqn.~\ref{eq:tangens}.

In the Aharonov-Bohm limit we have an analytical expression for
the cross section. It turns out that in this limit the integrand in
Eqn.~\ref{eq:txx} is constant. We have in this limit $\zeta =2\sin^2\pi f$,
or in other words
\begin{equation}
\frac{1}{\tau^{AB}}=n_v v_F \frac{\sin^2 \pi f}{\pi k}
=\omega_c \frac{\sin^2 \pi f}{\pi f},
\end{equation}
where $\omega_c$ is the cyclotron frequency corresponding to the flux density
of the external magnetic field.
We note that $\zeta$ vanishes in the Aharonov-Bohm limit,
when the flux fraction $f$ is integer, as it should.
We can use this expression to
make an estimate of the relative resistance change due to the vortices
$\Delta\rho_{xx}/\rho_{xx}=2 l_f n_v\sin^2\pi f / k_F $.
With $f=1/2$, $l_f=5\mu m$, $n=10^{11}{\rm cm^{-2}}$ and $B=100G$,
we get $\Delta\rho_{xx}/\rho_{xx}=0.6$, which is a significantly higher
value than is observed experimentally.
Let us now include the effects of the finite radius of the vortices.
Fig.~\ref{fig:zeta} shows several
$\zeta(\kappa)$ curves for different values of the flux fraction $f$. It is
seen that in general broader vortices give lower $\zeta$, i.e.\ less
resistance. The curve $f=1/2$
corresponds to the physical Abrikosov vortices, if the difference
in the cross sectional shape is ignored. In the
very low field limit, that is less than $100G$, the observed increase
in resistivity is linear in the applied field, and the relative change at
$B=100G$ is in the range $10^{-2}-10^{-3}$ \cite{Geim93},
much less than the present estimate gives.

The theory we have outlined
here is only valid when the vortices are well separated, this amounts to
assuming $d\gg R_v$. The crossover to a different kind of
behaviour observed in experiment, which is seen in both $\rho_{xx}$ and
$\rho_{xy}$, appear about $100G$, and we believe that this is the flux
density where the broad Abrikosov vortices begin to interfere.

\subsubsection{Transverse Resistivity}

The Hall resistivity, obtained from Eqn.~\ref{eq:current}, we can write as
\begin{equation}
\rho_{xy}=\alpha\frac{B}{ne},
\end{equation}
where $B$ is the externally applied homogeneous magnetic field. Here $B/ne$
is the Hall resistivity of a 2DEG in a homogeneous magnetic field, and
$\alpha$ is a dimensionless number, which describe the effects of the field
being inhomogeneous. The dimensionless quantity $\alpha$ is given by the
expression
\begin{equation}\label{eq:alpha}
\alpha=\frac{k_F}{2\pi f}
\int_{-\pi}^{\pi}d\theta\, \sin\theta |F(\theta)|^2
=\frac{\kappa}{2\pi f}\int_{-\pi}^{\pi}d\theta\, \sin\theta
\frac{d\varsigma}{d\theta}.
\end{equation}
Again, the integral can be expressed as a sum over terms involving only the
parameters $t_l=\tan\delta_l$, which are given by Eqn.~\ref{eq:tangens}
\begin{equation}
\alpha=\frac{2}{\pi f}\sum_{l=-\infty}^{\infty}
\frac{t_l t_{l+1} (t_{l+1} - t_l)}{(1+t_l^2)(1+t_{l+1}^2)}.
\end{equation}
Curves showing $\alpha$ as a function of $\kappa$ for different values of
the flux fraction $f$, have been plotted in Fig.~\ref{fig:alpha}.
The classical limit is realized when $R_v/\lambda_F\gg 1$ together with
$f\gg 1$, and we have $\kappa=2\pi\frac{R_v}{\lambda_F}$ so none of the curves
shown reach the classical regime.
The number of terms, which must be included in the sum over
$l$, grows rapidly with increasing $\kappa$ and $f$, thus making it difficult
to reach the true classical regime by this technique.

The $\alpha$-curve for $f=1/2$ we can compare with the experimental
Hall factor measured by Geim et al.\ \cite{Geim92}.
The overall qualitative behaviour is
in good agreement, when the very idealized shape of the vortices, we have used
in our calculation, is taken into account. To make a quantitative test we
have fitted our curve to the experimental curve by tuning the radius of the
vortex $R_v$. The best fit is obtained for a vortex radius $R_v=30$nm, and
this is nearly an order of magnitude smaller than the exponential length
estimated by Geim to be 100nm. This may indicate that most of the flux in the
vortices are concentrated in a narrow core.

It is seen in Fig.~\ref{fig:alpha} that the $\alpha$ curves corresponding
respectively to $f=1/4$, and $f=3/4$ does not seem to converge to the
level $\alpha=1$ as one would expect from the classical calculation.
This is another manifestation of the Aharonov-Bohm phenomenon. Consider the
ordinary double slit experiment with an infinitely thin solenoid hidden
behind the middle obstacle. The interference pattern which can be observed
behind the arrangement when it is hit by an incident plane wave, is
symmetric when the flux through the AB-solenoid is equal to zero,
modulo the flux quantum. In this case the interference pattern will have a
local maxima right in the middle. When the flux is equal to half a flux
quantum, modulo the flux quantum, the interference pattern will again be
symmetric, but this time with a node in the middle.
In both the cases $\phi=\phi_0$ and $\phi=\phi_0/2$ there are no net
scattering of electrons to either side.  But for general
fluxes $f\phi_0$, with $2f$ not equal to an integer, the interference
pattern is not symmetric, and this is the reason that the $f=1/4$ and $f=3/4$
$\alpha$-curves does not converge to the ``classical'' level at $\alpha=1$.
When $\kappa\gg 1$ and the scattering inside the vortex has become classically
behaived, the vortex still plays the role of an obstacle that gives rise to
an interference pattern, and the vector potential {\em outside} the vortex
deflects the pattern and thereby gives rise to the asymmetry we observe in
Fig.~\ref{fig:alpha}. We emphasize that in the limit where the diameter of the
vortex (the obstacle)
goes to zero, there is no net scattering to either side.

\subsection{Multi Flux Quantum Vortex and Resonance \- Scattering}
\label{sec:reso}

When the total amount of magnetic flux inside the vortex is increased, the
$\alpha$ and $\zeta$ spectra acquire more structure. In Fig.~\ref{fig:reso} we
have shown $\alpha$ and $\zeta$ curves for a flux vortex carrying a total of
10 flux quanta.
The structure seen in the plots is an effect of the resonant scattering
which takes place when the energy of the incoming particle is
close to one of the Landau quantization energies corresponding to the
magnetic field strength inside the vortex. The magnetic field in the vortex
vanishes outside a finite range -- the radius of the vortex -- and there
are therefore no real Landau levels in the sense of stationary eigenstates,
but only metastable states.
In the dimensionless units we are working with, the Landau quantization
energies $E_p=\hbar\omega_c(p+1/2)$ corresponds to
\begin{equation}
\kappa_p=2\sqrt{|f|(p+\frac{1}{2})},\hspace{0.5cm}p=0,1,2,\dots
\end{equation}
These values are in excellent agreement with the resonances seen in
Fig.~\ref{fig:reso}, where the first eight resonances corresponding to
$p=0,\dots,7$, are clearly distinguished.
At the resonance energies the typical time the particle spends
in the scattering region, i.e.\ inside the vortex, is much longer than it
is away from the resonance. The time the particle spends inside the vortex at
a resonance, can be thought of as the lifetime of the corresponding metastable
state. The inverse lifetime is proportional to the width of the resonance,
that is strictly speaking the width of the peak in the partial wave cross
section $\sigma_l$, corresponding to the $l$ quantum number of the
metastable Landau state.

It is easy to interpret the small peaks in the $\zeta$-curve in
Fig.~\ref{fig:reso}, appearing at the resonance energies. Because when the
electron spends longer time in the scattering region, it loses knowledge of
where it came from, resulting in an enhanced probability of being scattered
in the backwards direction. The $\alpha$-curve is a measure of asymmetric
scattering, and we can therefore interpret the dips seen in
Fig.~\ref{fig:reso} along the same line of reasoning as for the
$\zeta$-peaks. The electron spends longer time in the scattering region,
thereby losing knowledge of what is left and what is right. To explain why
the $\alpha$-curve is asymmetric in $\kappa$ around the
resonances one could look at it this way: For increasing $\kappa$ the
electron scattering becomes more and more classical, giving rise to the
overall increasing background in $\alpha$. But every time a new scattering
channel is opened, the asymmetry of the scattering is suppressed, due to the
lack of knowledge effect, thus resulting in a sawtooth like curve.

\section{Hall Effect in a Regular Array of Flux Vortices}\label{sec:array}

\subsection{Introduction}
\label{sec:intro}

Recently measurements were made by Geim et al.\ \cite{Geim92,Geim93} of the
Hall resisti\-vi\-ty of low density 2DEG's in a random distribution of flux
vortices, at very low magnetic field strengths. A profound suppression
of the Hall resistivity was found, for 2DEG's with Fermi wavelengths
of the same order of magnitude as the diameter of the flux vortices.
This indicates that we are dealing with a phenomenon of quantum nature.
These measurements were made by placing a thin lead film on top of a
GaAs/GaAlAs heterostructure.
When a perpendicular magnetic field is applied,
the magnetic field penetrates the superconducting lead film and also
the heterostructure in the form of flux vortices
each carrying half a flux quantum $\phi_0 /2$ of magnetic flux. Due to the
strong flux vortex pinning in the films Geim have used, the vortices
were positioned in a random configuration.

In this section we will consider the hypothetical experiment where one
instead of a ``dirty'' film, places a perfectly homogeneous type II
superconducting film, on top of the 2DEG.
The film do not have to be made of a material
which is type II superconducting in bulk form. A film of a type I
superconducting material will also display a mixed state if the thickness of
the film is below the critical thickness $d_c$. Experimentally perfect
Abrikosov flux vortex lattices have been observed in thin films of lead
with thickness $d<d_c\cong 0.1\mu$m, \cite{Hyb}.

If one succeeds to make such a sandwich construction, one has an ideal
system for investigating how a 2 dimensional electron gas behaves in a
periodic magnetic field. When the magnetic field exeeds $H_{c1}$, which can be
extremely low, the superconducting film will enter the mixed phase, and
form an Abrikosov lattice of flux vortices. The
Abrikosov lattice in the superconducter will give rise to a periodic magnetic
field at the 2DEG, and moreover as the strength of the applied magnetic field
is varied the only difference at the 2DEG, is that the lattice constant of
the periodic magnetic field varies. The Abrikosov lattice is most often a
triangular lattice with hexagonal symmetry, although other lattices have
been observed (e.g.\ square) in special cases where the atomic lattice
structure impose a symmetry on the flux lattice, \cite{Hyb}.
For simplicity the model
calculations which we have done, were made for a system with a square lattice
of flux vortices, but we do not expect this to influence the overall
features of the results. From the point of view of the 2DEG it is
important that the flux vortices carry half a flux quantum
$\frac{\phi_0}{2} = \frac{h}{2e}$ due to the $2e$-charge of the Cooper pairs
in the superconductor. The magnetic field from a single flux vortex fall of
exponentially with the distance from the center of the vortex. This
exponential decay is characterised by a length $\lambda_s$, which essentially
is the London length of the superconducter, proportional to one over the
square root of the density of Cooper pairs. The length $\lambda_s$ can be
varied by changing the temperature, or the material of the superconductor.

The other characteristic lengths of the system are the Fermi wavelength
$\lambda_F =\sqrt{\frac{2\pi}{n}}$, where $n$ is the density of the 2DEG,
the lattice constant $a$ of the periodic magnetic field, and the mean free
path $l_f=v_F \tau$. The mean free path we assume to be very large compared to
$a$ and $\lambda_s$. The magnetic field is only varying appreciably when $a$
is larger than $\lambda_s$. This means that the magnetic flux density of the
applied field should be appreciably less than $\phi_0/(\pi\lambda_s^2)$,
which typically is of order 1000Gauss. In the limit where
$\lambda_s \ll a,\lambda_F$, the vortices
can be considered magnetic strings, and the electrons experiences a periodic
array of Aharonov-Bohm scatterers.
In this case the value of the flux through each
vortex is crucial. If for instance the flux had been one flux quantum
$\phi_0=\frac{h}{e}$, the electrons would not have been able to feel the
vortices at all. But in the real world the vortices from the superconductor
carry $\frac{\phi_0}{2} = \frac{h}{2e}$  of flux, and therefore this limit is
nontrivial. The electrons has for instance a band structure quite different
from that of free electrons. In the mathematical limit of infinitely thin
vortices each carrying {\em half} a flux quantum, there cannot be any
Hall effect. This is most easily seen by subtracting one flux quantum
from each vortex to obtain a flux equal to minus half a flux quantum
through each vortex. As we have discussed earlier the introduction of the
Dirac strings can not change any physics, and the procedure therefore shows
that the system is equivalent to it's time reversed counterpart, thereby
eliminating the possibility of a Hall effect.

In this study we have ignored the electron spin throughout, in order to keep
the model simple. From the point of view of the phenomena we are going to
describe, the effect of the electron spin will be to add various small
corrections.

\subsection{Electrons in a periodic magnetic field}
\label{sec:elec}

\subsubsection{Magnetic translations}

It is a general result for a charged particle in a spatially periodic
magnetic field $B(x,y)$, that the eigenstates of the system can be labeled by
Bloch vectors taken from a Brillouin zone, if and only if the flux through
the unit cell of the magnetic field is a rational number $p/q$ times the
flux quantum.
The standard argument for this fact is made by introducing magnetic
translation operators.
To introduce magnetic translation operators in an inhomogeneous
magnetic field we first make the following observation.
The periodicity of the field can be stated $B(\bbox{r}+\bbox{R})=B(\bbox{r})$,
for $\bbox{R}$ belonging to a Bravais lattice. But this implies that the
difference between the vector potentials $\bbox{A}(\bbox{r}+\bbox{R})$ and
$\bbox{A}(\bbox{r})$ must be a gauge transformation
\begin{equation}
\nabla\times\left\{\bbox{A}(\bbox{r}+\bbox{R}) - \bbox{A}(\bbox{r})\right\}=
B(\bbox{r}+\bbox{R}) - B(\bbox{r}) =0.
\end{equation}
We introduce the gauge potential $\chi_{R}$ and write
\begin{equation}
\bbox{A}(\bbox{r}+\bbox{R}) = \bbox{A}(\bbox{r}) +\nabla\chi_{R}(\bbox{r}).
\end{equation}
The function $\chi_{R}$ is only defined modulo an arbitrary additive constant
which have no physical effect.
The Hamiltonian of the electrons is
\begin{equation}\label{eq:ham}
H = \frac{1}{2m}\left(\bbox{p} + e\bbox{A}(\bbox{r})\right)^{2}.
\end{equation}
The ordinary translation operators
$T_{R}=\exp[\frac{i}{\hbar}\bbox{R}\cdot\bbox{p}]$ do not commute with the
Hamiltonian, because they shift the
argument of the vector potential from $\bbox{r}$ to $\bbox{r}+\bbox{R}$, but as
we just have seen this can be undone with a gauge transformation. We therefore
introduce the magnetic translation operators as the combined symmetry
operation of an ordinary translation and a gauge transformation
\begin{equation}
M_{R}=\exp[-i\frac{e}{\hbar}\chi_{R}(\bbox{r})]
      \exp[\frac{i}{\hbar}\bbox{R}\cdot\bbox{p}].
\end{equation}
The operator $M_{R}$ is unitary, as it is the product of two unitary
operators, and therefore has eigenvalues of the form $e^{i\lambda}$. Let us
denote the primitive vectors of the Bravais lattice $\bbox{a}$ and $\bbox{b}$.
We can find common eigenstates of $M_{a}$, $M_{b}$ and $H$,
if and only if they all commute with each other. The magnetic translations
each commute with the Hamiltonian by construction, and furthermore we have
\begin{equation}
M_{a}M_{b} =  \exp[2\pi i\frac{\phi}{\phi_0}] M_{b}M_{a}.
\end{equation}
If the flux $\phi$ through
the unit cell is a rational number $p/q$ ($p$ and $q$ relatively prime) times
the flux quantum $\phi_0$, $M_{qa}$ and $M_{b}$ commute. In this
case the cell spanned by $q\bbox{a}$ and $\bbox{b}$ is called the magnetic
unit cell. Let us define $\bbox{c}=q\bbox{a}$. The possible eigenvalues of
$M_{c}$ are phases $e^{2\pi i k_1 }$, where we can restrict
$|k_1 |<\frac{1}{2}$, and equivalently for $M_{b}$. We can therefore
label the common eigenstates
$|\bbox{k},n\rangle$, where $\bbox{k}=k_1 \bbox{c}^* + k_2\bbox{b}^*$, and
$\bbox{c}^*,\bbox{b}^*$ are the primitive vectors of the reciprocal lattice.
The vector $\bbox{k}$ is restricted to the magnetic Brillouin zone.
An arbitrary magnetic translation of an eigenstate with a Bravais lattice
vector $\bbox{R}=n\bbox{c}+m\bbox{b}$ can now be written
$M_{R}|\bbox{k},n\rangle=(M_{c})^n (M_{b})^m |\bbox{k},n\rangle
=\exp[i \bbox{k}\cdot\bbox{R}]|\bbox{k},n\rangle$, showing that the eigenstate
is a Bloch state. In this case we can speak of
energy bands forming a band structure in the usual sense. When the flux
through the elementary unit
cell $(\bbox{a},\bbox{b})$ is an irrational number of flux quanta, the
situation
is different. The irrational number can be reached as the limit where $p$ and
$q$ get very large, and consequently the Brillouin zone get very small and
collapses in the limit.

\subsubsection{The Dirac vortex viewpoint}

In this section we will show, how it is possible to argue in a slightly
different way from the previous section, and hereby in a simpler way  obtain
the vector potential of a periodic magnetic field.
 Let us again assume a rectangular unit cell
$(a,b)$, $B(x+a,y)=B(x,y+b)=B(x,y)$ etc., to keep the notation simple.
The magnetic field enter the Hamiltonian only through the vector potential.
The question is therefore if one can choose a gauge such that the
vector potential will be translationnally invariant relative to a unit cell
$(c,d)$ $\bbox{A}(x+c,y)=\bbox{A}(x,y+d)=\bbox{A}(x,y)$ etc. It is clear that
if
such a periodic $\bbox{A}$-field exists, then the total flux $\Phi_{cd}$
through the unit cell $(c,d)$ will be zero, as it is given by the line
integral of $\bbox{A}$ around the boundary of the unit cell $(c,d)$
\begin{equation}
\Phi_{cd}=\oint_{\partial (c,d)} \bbox{A}\cdot\bbox{dl},
\end{equation}
which is zero by the periodicity.
We remark that due to the relation $B=\nabla\times\bbox{A}$, the cell $(c,d)$
will be bigger than or equal to $(a,b)$. If the flux
through the unit cell of the magnetic field is not zero, but equal to a
rational number times the flux quantum, $\Phi_{ab}=p/q\;\phi_0$ ($p$ and $q$
relatively prime), a trick can be applied to make the flux $\Phi_{cd}$
become zero.
It is a basic fact, apparently first observed by Dirac~\cite{Dirac},
that a particle with charge $e$ cannot feel an infinitely thin solenoid
carrying a flux equal to an integer multiple of the flux quantum
$\phi_{0}=h/e$.
Such a stringlike object carrying one flux quantum is sometimes called a
Dirac vortex.
On a lattice the Dirac vortex goes through the center of a plaquette,
and the electron can therefore never enter the core of it. The lattice Dirac
vortex can be made to disappear by a gauge transformation.

To find the periodic vector potential take an enlarged unit cell
$(c,d)=(qa,b)$, so that $\Phi_{cd}=p\phi_0$ and
put by hand $p$ counter Dirac vortices through the cell, to
obtain zero net flux. Then a divergence free vector potential can be build for
instance by Fourier transform
\begin{eqnarray}\label{eq:FourierA}
B(\bbox{Q})=\frac{1}{cd}\int_{(c,d)}d^2 r \exp[-i\bbox{Q}\cdot\bbox{r}]
B(\bbox{r}), \\
A(\bbox{r})=\sum_{Q\neq 0}\left(\begin{array}{c}
iQ_y \\ -iQ_x \end{array}\right)\frac{\exp[i\bbox{Q}\cdot\bbox{r}]}{Q^2}
B(\bbox{Q}),\label{eq:FourierB}
\end{eqnarray}
where the sum is over  $\bbox{Q}$ in the reciprocal lattice.
Here we have used continuum notation, but it is straightforward to write
down the lattice equivalents of the expressions.

\subsection{Lattice calculation of Hall conductivity}

We have calculated the Hall conductance of the 2DEG in the vortex field by a
numerical lattice method. This we do because the calculations then reduces to
linear algebra operations on finite size matrices, which can be implemented
in a C++ program on a computer.
The idea is to consider an electron moving on a
discrete lattice, rather than in continuum space. We know, allthough we are
not going to prove it here, that in the limit where the discrete lattice
becomes finegrained compared to all other characteristic length of the system,
the continuum theory is recovered. Here we assume that the original Bravais
lattice has square lattice symmetry, with a lattice parameter which we
call~$a$. The discrete micro lattice is then introduced as a finegrained
square lattice inside the unit cell of the Bravais lattice. The lattice
parameter of the micro lattice we then take as $a/d$, where $d$ is some
large number, in order to keep the to lattices commensurable. The condition
that we have to impose on the micro lattice, in order that it is a good
aproximation to the continuum, can then be stated
\begin{equation}
a/d\ll a,\lambda_F, \lambda_s,\cdots .
\end{equation}
In the numerical calculations we have made, we have taken $d=10$.

The tight-binding calculations are made with the Hamiltonian
\begin{equation}
H = -\sum_{ij\tau\tau'}t_{i+\tau,j+\tau'}c^{\dag}_{i+\tau}c_{j+\tau'}.
\end{equation}
Here $\bbox{i},\bbox{j}$ are Bravais lattice vectors, and
$\bbox{\tau},\bbox{\tau}'$
are vectors indicating
the sites in the basis. The matrix elements $t_{i+\tau,j+\tau'}$ are
taken non-zero only between nearest neighbour sites.
The matrix element between
two nearest neighbour sites $\tau$ and $\tau'$ are complex variables
$t_{\tau,\tau+e_{\mu}}=t e^{iA_{\mu}(\tau)}$ with a phase
given by the vector potential $A_{\mu}(\tau)$ residing on the link joining
the sites.
The translation invariance of the Hamiltonian can
then be stated $t_{i+\tau+l,j+\tau'+l} = t_{i+\tau,j+\tau'}$, for all vectors
$\bbox{l}$ belonging to the Bravais lattice.
Let us introduce the system on which our
calculations were made as an example. Fig.~\ref{basis} shows the unit cell
with its internal structure i.e. the basis.
There are $N=d\cdot d$ sites
in the basis. The length of the links we write as $a/d$, where $a$ is the
side of the unit cell, with area $\Omega = a^2$.
The vectors $\bbox{\tau}= (\tau_1,\tau_2)\frac{a}{d},$
$\tau_1,\tau_2 =0,1,\dots d-1$
are offsets into the basis, while the vectors $\bbox{i}=(i_1,i_2)a,$
$i_1,i_2 \in \Bbl{Z}$ indicate the cells in the Bravais lattice.
The operator $c^{\dag}_{i+\tau}$, for
a given $\tau$, is defined on the Bravais lattice, and accordingly it
can be resolved as a Fourier integral over the Brillouin zone as
\begin{equation}
c^{\dag}_{j+\tau} = \frac{\Omega}{(2\pi)^2}\int_{BZ} d^2 q
e^{-i\bbox{q} \cdot (\bbox{j} + \bbox{\tau})}c^{\dag}_{q,\tau}.
\end{equation}
(It should be noted that the factor $e^{-i\bbox{q}\cdot\bbox{\tau}}$ is
arbitrary and included here for later convenience).
Inserting this and using the translation invariance, the Hamiltonian can be
rewritten as
\begin{equation}
H = \frac{\Omega}{(2\pi)^2}\int_{BZ}d^2 k H_{k},
\end{equation}
where we have introduced
\begin{equation}
H_{k} = -\sum_{j,\tau,\tau'}t_{\tau,j+\tau'}e^{i\bbox{k}\cdot(\bbox{j}
+ \bbox{\tau}'-\bbox{\tau})}c^{\dag}_{k,\tau}c_{k,\tau'}.
\end{equation}
It is seen that $H_{k}$ only mixes the $N$ states $|\bbox{k}\tau
\rangle$, i.e.\ it is an $N\times N$ matrix. The $N$ eigenvalues of
$H_{k}$ are the energies of
the $N$ tight binding Bloch states with wavevector $\bbox{k}$. Let us denote
the eigenstates of $H_{k}$ by $u^{\alpha}_{k}$
\begin{equation}\label{eq:schrodinger}
H_{k} u^{\alpha}_{k} = E^{\alpha}_{k} u^{\alpha}_{k}
\end{equation}
where $\alpha=1,2,\dots N$ and $E^{\alpha}_{k} \le E^{\alpha +1}_{k}$.
{}From the $N$ dimensional vector $u^{\alpha}_{k}$ we can construct the
eigenstate $\Psi^{\alpha}_{k}$ of the Hamiltonian $H$
\begin{equation}
\left\langle j+\tau |\Psi^{\alpha}_{k}\right\rangle =
e^{i\bbox{k}\cdot(\bbox{j}+\bbox{\tau})} u^{\alpha}_{k}(\tau).
\end{equation}
It is straightforward to verify that this is the correct Bloch eigenstate of
$H$. The bandstructure can be calculated directly by diagonalising the
$N\times N$ matrices, $H_{k}$, for representative choices of $\bbox{k}$ in
the Brillouin zone. Before one can compare the spectrum optained from this
calculation with that of a continuum system, a scaling of the energies is
required. To scale the energy to the spectrum of a particle with an effective
mass $m$, we have to take $t=\hbar^2 d^2 /ma^2$,
and $\epsilon_{\alpha}(\bbox{k})=E^{\alpha}_k + 4t$.

The Hall conductivity can be calculated by the same method as in the
homogeneous magnetic field~\cite{TKNN82,Kohmoto85}.
We have used a single particle Kubo formula to calculate the Hall
conductance
\begin{equation}
\label{eq:sigmasum}
\sigma_{xy}=\frac{i \hbar}{A_0}\sum_{E^{\alpha}<E_F <E^{\beta}}
\frac{(J_x)_{\alpha\beta}(J_y)_{\beta\alpha} - (J_y)_{\alpha\beta}
(J_x)_{\beta\alpha}}{(E^{\alpha} - E^{\beta})^2},
\end{equation}
where $J_x$,$J_y$ are the currents in the $x$,$y$ directions, and the sum is
over single particle states $|\alpha,\bbox{k}\rangle$ with energies below and
above the Fermi level $E_F$. The area of the system is denoted $A_0$.
All quantities are diagonal in $\bbox{k}$, and
therefore this index is suppressed. The summation is composed of a discrete
sum over bands, and an integral over the Brillouin zone for each band. The
Brillouin zone shown in Fig.~\ref{Brillouin} is doubly connected because the
states on the edges is to be identified according to the translation
invariance.
This gives the Brillouin zone the topology of a torus $T^2$, with
two basic non contractible loops.
The current operator can be written
\begin{equation}
\bbox{J}=\frac{\Omega}{(2\pi)^2}\int_{BZ}d^2 k \bbox{J}_k,
\end{equation}
where $\bbox{J}_k = \frac{e}{\hbar}\frac{\partial H_k}{\partial\bbox{k}}$.
By use of some simple manipulations and completeness, it is straightforward
to refrase Eqn.~\ref{eq:sigmasum}
\begin{equation}\label{eq:sigma}
\sigma_{xy}=\frac{i e^2}{\hbar A_0}\sum_{E^{\alpha}<E_F}\left(
\left\langle\frac{\partial\alpha}{\partial k_x}\left|
\frac{\partial\alpha}{\partial k_y}\right.\right\rangle -
\left\langle\frac{\partial\alpha}{\partial k_y}\left|
\frac{\partial\alpha}{\partial k_x}\right.\right\rangle
\right),
\end{equation}
where $\left|\frac{\partial\beta}{\partial k_{\mu}}\right\rangle$ is
shorthand for $\frac{\partial}{\partial k_{\mu}}\left| \beta,\bbox{k}
\right\rangle$.
This formula was first derived by Thouless, Kohmoto, Nightingale and
den Nijs \cite{TKNN82},
for a noninteracting 2 dimensional electron gas in a periodic scalar
potential, and a commensurate perpendicular magnetic field. It requires some
comments to be meaningful. In order to calculate $\left|\frac{\partial
\alpha}{\partial k_{\mu}}\right\rangle$ it is necessary to consider the
difference $(\left|\alpha,\bbox{k}+\delta\bbox{k_{\mu}}\right\rangle -
\left|\alpha,\bbox{k}\right\rangle)/\delta k_{\mu}$. But this difference is not
well defined as it stands, as the {\sl phase} of the states is arbitrary.
Rather than representing the state $u_{k}$ by a
single vector in $\Bbl{C}^N$, it should be represented by a class of vectors
which differ one from another only by a phase. These equivalence classes are
sometimes called rays. To compare states locally, we need to project this
$U(1)$ degree of freedom out. This is done by demanding the wave function
to be real, when evaluated in a fixed point, i.e.\ $u^{\alpha}_k (\tau_i)=
\langle\tau_i |\alpha,\bbox{k}\rangle \in \Bbl{R}$. If the wave function
happens to be zero in $\tau_i$, some other point $\tau_j$ must be used.
When a band has a non-zero Hall conductivity, it is not possible to find
a single $\tau$ which work for all the states in the Brillouin zone. The
change from $\tau_i$ to $\tau_j$ which shifts the phase of the states,
is analogous to a gauge transformation on the set of states. The special
combination of terms which appear in Eqn.~\ref{eq:sigmasum} is gauge
invariant with respect to these special ``gauge transformations''.
If we let $|\chi^{\alpha}\rangle$ denote a state which is obtained from
$|\alpha,\bbox{k}\rangle$ by fixing the phase according to the above scheme,
the following formula for the contribution to the Hall
conductivity from a single band $\alpha$, is well defined
\begin{equation}\label{eq:sigmawelldef}
\sigma^{\alpha}_{xy}=\frac{e^2}{h}\frac{1}{2\pi i}\int_{BZ}d^2 k
\left\{\left\langle\frac{\partial\chi^{\alpha}}{\partial k_y}\left|
\frac{\partial\chi^{\alpha}}{\partial k_x}\right.\right\rangle -
\left\langle\frac{\partial\chi^{\alpha}}{\partial k_x}\left|
\frac{\partial\chi^{\alpha}}{\partial k_y}\right.\right\rangle
\right\}.
\end{equation}
  It has been shown in detail by Kohmoto \cite{Kohmoto85}, for the homogeneous
magnetic field case, that this expression is equal to minus $\frac{e^2}{h}$
times the first Chern
number of a principal fibre bundle over the torus.
As the first Chern number is always an integer, this has the physical
consequence that whenever the Fermi energy lies in an energy gap, the
Hall conductance is quantized.
We will use this result to
interpret certain peaks in the $\sigma_{xy}$-spectra we have calculated.
When the Fermi level is not in an energy gap of the system, we will have to
use Eqn.~\ref{eq:sigma} to calculate $\sigma_{xy}$. As we shall see, in this
case there is no topological quantization of the Hall conductivity.

\subsection{Energy band crossing}\label{sec:bandcross}

In this section we study the effect on the Hall conductivity of an energy
band crossing. This has previously been discussed in different contexts by
several authors \cite{Berry84,Wilk84,Axel}.

When the shape of the magnetic field is varied, controlled by some outer
parameter $\xi$, it will happen for certain parameter values $\xi_0$, that two
bands cross, see Fig.~\ref{bandcross}.
This is the consequense of the
Wigner-von Neumann theorem, which states that three parameters are required in
the Hamiltonian in order to produce a degeneracy not related to symmetry.
Here the parameters are
$k_x,k_y$ and the outer parameter $\xi$, which in our calculation is the
exponential length of the
flux vortices from the superconducter. When the energy difference $E^+ - E^-$
between the two bands considered is much smaller than the energy distance to
the other bands, the Hamiltonian can be restricted to the subspace spanned by
the two states $|+,\bbox{k}^0\rangle$ and $|-,\bbox{k}^0\rangle$.
The point in the Brillouin zone where the degeneracy occur we
denote $\bbox{k}^0$. The Hamiltomian $H(\bbox{k})$ is
diagonal for $\bbox{k}=\bbox{k}^0$, and we denote the diagonal elements
respectively $E_0 + \epsilon$ and $E_0 - \epsilon$.
For small deviations of $\bbox{k}$ from $\bbox{k}^0$ the
lowest order corrections to the Hamiltonian is offdiagonal elements
$\Delta(\bbox{k})$ linear in $\bbox{k}-\bbox{k}^0$.
Without essential loss of generality we can assume that $\epsilon$ is
independent of $\bbox{k}$. Then $\epsilon$ plays the role of the outer
parameter controlling the band crossing. The Hamiltonian is then
approximated by
\begin{equation}\label{eq:apx}
H(\bbox{k})=\left(
\begin{array}{cc}
\epsilon & \Delta^* \\
\Delta & -\epsilon
\end{array}
\right) + E_0.
\end{equation}
The offdiagonal element is expanded as
\begin{equation}
\Delta(\bbox{k})=\alpha(k_x -k^0_x) + \beta(k_y -k^0_y)
\end{equation}
with $\alpha=\frac{\partial}{\partial k_x}\langle -,\bbox{k}^0|
H_{k}|+,\bbox{k}^0\rangle$, and
$\beta=\frac{\partial}{\partial k_y}\langle -,\bbox{k}^0|
H_{k}|+,\bbox{k}^0\rangle$.

We want to find the consequenses of the energy band degeneracy, on the
topological Hall quantum numbers of the bands.
Let us define
\begin{equation}
B_{\pm}(\bbox{k})=
\left\{\left\langle\frac{\partial\pm}{\partial k_y}\left|
\frac{\partial\pm}{\partial k_x}\right.\right\rangle -
\left\langle\frac{\partial\pm}{\partial k_x}\left|
\frac{\partial\pm}{\partial k_y}\right.\right\rangle
\right\}.
\end{equation}
Then the  interesting quantities are
the integrals of $B_{+}(\bbox{k})$ and $B_{-}(\bbox{k})$, around a small
neighbourhood of the degeneracy point $\bbox{k}^0$.
It turns out that it is the two numbers $\alpha$ and $\beta$ that control
what happens.

In general $\alpha$ and $\beta$ will be nonzero complex numbers --- nonzero
because we have assumed the degeneracy to be of first order. Let us first
consider the degenerate case where $\alpha$ and $\beta$ are linearly
dependent, i.e.\ $\alpha/\beta$ is real, or otherwise stated
Im$(\alpha^* \beta)=0$. Then by a linear transformation we can write the
Hamiltonian
\begin{equation}
h(\bbox{\kappa})=(\kappa_1+\kappa_2) \sigma^1 + \gamma\sigma^3 =
\left(
\begin{array}{cc}
\gamma  &  \kappa_1 +\kappa_2  \\
\kappa_1 + \kappa_2 & -\gamma
\end{array}\right),
\end{equation}
where now $\gamma$ is the dimensionless parameter of the crossing,
and $\kappa_1,\kappa_2$ are the rescaled dimensionless momentum variables.
The $\sigma^{\mu}$'s refers to the Pauli matrices.
This Hamiltonian is real, and we can therefore also choose the eigenstates
to be real, and this will clearly lead to a vanishing $B(\bbox{\kappa})$. In
this case we therefore conclude that there is no exchange of topological
charge. Here the word topological charge is used to denote Hall quanta.

Let us now treat the general case where $\alpha$ and $\beta$ are linearly
independent, i.e.\ Im$(\alpha^* \beta)\neq 0$. In this case, we can by a
linear transformation write
$\Delta(\bbox{k})/E_0=\kappa_1 + i\kappa_2=\kappa e^{i\theta}$,
which defines the scaled momentum variables $\kappa,\theta$.
(Here $E_0$ is some constant with dimension of energy.)
This reduces the Hamiltonian to the form
\begin{equation}
h(\kappa,\theta)=\kappa_1 \sigma^1 + \kappa_2 \sigma^2 + \gamma\sigma^3 =
\left(
\begin{array}{cc}
\gamma  &  \kappa e^{-i\theta} \\
\kappa e^{i\theta} & -\gamma
\end{array}\right).
\end{equation}
Let us define $\lambda=\sqrt{\gamma^2 + \kappa^2}$. Then the eigenvalues
of $h(\kappa,\theta)$ are $\pm\lambda$ and the two corresponding eigenstates
are
\begin{equation}
\left|\pm,\bbox{k}\right\rangle=\frac{1}{\sqrt{2}}
\left(\begin{array}{c}
\pm\sqrt{1\pm \gamma/\lambda} \\ e^{i\theta}\sqrt{1\mp \gamma/\lambda}
\end{array}\right).
\end{equation}
In order to calculate the integral of the $B$-function we need
to express it in terms of the $\kappa,\theta$-variables. The Jacobian of the
transformation is given by the expression
\begin{equation}
dk_x dk_y = \frac{\kappa\, d\kappa d\theta}{|\alpha_r\beta_i -
\alpha_i\beta_r|},
\end{equation}
and
\begin{equation}
B_{\pm}(\kappa,\theta)=\frac{\alpha_r\beta_i - \alpha_i\beta_r}{\kappa}
\left\{\left\langle\frac{\partial\pm}{\partial \theta}\left|
\frac{\partial\pm}{\partial \kappa}\right.\right\rangle -
\left\langle\frac{\partial\pm}{\partial \kappa}\left|
\frac{\partial\pm}{\partial \theta}\right.\right\rangle
\right\}
=\pm (\alpha_r\beta_i - \alpha_i\beta_r)\frac{i\gamma}{2\lambda^3},
\end{equation}
where the indices $r,i$ refer to the real and imaginary parts respectively.
We can now calculate the contribution to the Hall conductivity from each of
the bands, from the area around $\bbox{k}^0$ given by
$|\kappa |<\kappa_c$, where $\kappa_c$ is some local
cutoff parameter which limit the integration to the area where the
approximation leading to the Hamiltonian Eqn.~\ref{eq:apx} is valid
\begin{eqnarray}
\Delta\sigma^{\pm}_{xy} & = & \frac{e^2}{h}\frac{1}{2\pi i}
\int dk_x \int dk_y B(\bbox{k}) \nonumber \\
& = & \frac{e^2}{h}\frac{1}{2\pi i}{\rm sign}[{\rm Im} (\alpha^* \beta)]
\int_0^{\kappa_c} d\kappa \int_0^{2\pi} d\theta
\frac{i\kappa\gamma}{2\lambda^3} \nonumber \\
& = & \frac{e^2}{2h}{\rm sign}[\alpha^* \beta ] \frac{\gamma}{|\gamma|}
\int_0^{\kappa_c/|\gamma|}\frac{u\,du}{(1+u^2)^{3/2}} \nonumber \\
& = & \pm\frac{e^2}{2h}
{\rm sign}[{\rm Im} (\alpha^* \beta)] {\rm sign}[\gamma],
\end{eqnarray}
where the last equality sign is valid when
$\kappa_c/|\gamma|\gg 1$, i.e.\ close to the crossing where
$\gamma=0$.
Here the factor sign$[\gamma]$ signals that the two bands exchange exactly one
topological conductivity
quantum $\frac{e^2}{h}$, at the crossing. This is not surprising, because
we know that the total contribution from the states in a single band is
always an integer times the quantum $\frac{e^2}{h}$. Moreover it is readily
seen that in the hypothetical situation of a $n$'th order degeneracy,
i.e.\ one for which $\Delta=\kappa^n e^{i n \theta}$, $n$ quanta are
exchanged.
The total topological charge in the
band structure is conserved. The topological charge can flow around and
rearrange itself inside a band, but only be exchanged between bands in lumps
equal to an integer multiplum of the conductivity quantum
$\frac{e^2}{h}$. When the radius of the flux vortices is
gradually shrinked to zero, the Hall effect has to disappear.
There are two mechanisms with which the Hall
effect can be eliminated. The first is by moving the topological charge up
through the band structure by exchanging quanta, resulting in a net upward
current of topological charge, eventually moving the charge up above the Fermi
surface,
where it has no effect. The second mechanism is by rearranging the topological
charge inside the bands, so that each band has a large negative charge in the
bottom, and a large positive charge in the top, but arranged in such a clever
way that charge neutrality is more or less retained for all energies. This
second mechanism will also give a net displacement of topological charge up
above the Fermi energi, because in general the Fermi surface cuts a great many
bands, and for all these bands the large negative charge, which they have
in their bottom part, will be uncompensated. To use the language of
electricity theory, we
can say that every band gets extremely polarized, resulting in a net upward
displacement current, in analogy with the situation in a strongly polarized
dielectric. Our numerical calculations indicate, that it is the second
mechanism which is responsible for the elimination of the Hall effect,
as the radius of the vortices shrinks to zero.

When two bands are nearly degenerate for some $\bbox{k}^0$,
each of the bands have concentrated  half a quantum in a small area in k-space
around $\bbox{k}^0$, and in general the topological charge piles up across
local and global gaps in the energy spectrum.
This is the reason for the oscillatory and spiky behaviour
of the Hall conductance as a function of electron density, that is seen on the
calculated spectra below. It is also the reason why the numerical integration
involved in the actual evaluation of the Hall conductivity, is more tricky
than one could wish.
In particular it indicates that it is not true as sometimes
conjectured, e.g.\ \cite{Pffan},
that the Hall conductivity is smoothly distributed in
k-space, and that it therefore should vary smoothly between the quantized
values at the energy gaps, as the Fermi energy is swept through a band.

\subsection{Numerical results}\label{sec:num}
\subsubsection{Transverse Conductivity}

We have calculated the transverse conductivity $\sigma_{xy}$ as a function of
the integrated density of states, for electrons in a square lattice of flux
vortices, for a series of varying cross sectional shapes of the
flux vortices.
Each of the field configurations consists of a square lattice
of flux vortices with a given exponential length~$\lambda_s$. The parameter
which vary from calculation to calculation, is the dimensionless
ratio $\xi=\lambda_s /a$, where $a$ is the length of the edges of the
quadratic unit cell. The unit cell is shown
on Fig.~\ref{basis} and contain, as we have already discussed, two vortices
and a counter Dirac vortex. In
order to do the tight binding calculation a micro lattice is introduced in the
unit cell. In all the numerical calculations we present,
the micro lattice is $10\times 10$. This gives 100 energy bands
distributed symmetrically about the center on an energy scale. Out of these
only the lower part, say band 1 to 20, approximate the
real energy bands well, while the rest is significantly affected by the finite
size of the micro lattice. A careful examination of the vector potential
reveal that the symmetry of the Hamiltonian is very high for the particular
choice of unit cell shown in Fig.~\ref{basis}. The field from a single vortex
we have taken as $B_0 e^{-(|\tau_x |+|\tau_y |)/\lambda_s}$ instead of the
more realistic $B_0 e^{-|\bbox{\tau}|/\lambda_s}$. With this choice the
vector potential can be written down analytically in closed form.
This makes the calculations
simpler, and does not break any symmetry that is not already broken by the
introduction of the micro lattice. (We have made calculations of
the band structure, with both kinds of flux vortices, and the differences are
indeed very small).
The energy spectrum is invariant under
the changes $(k_x,k_y)\longmapsto (\pm k_x,\pm k_y),(\pm k_y,\pm k_x)$. This
fact is exploited to present the band structures in an economic way. The
labels $\Gamma,X$ and $M$ correspond to the indicated points in the Brillouin
zone, Fig.~\ref{Brillouin}.

In Fig.~\ref{fig:halfbands} we have plotted a selection of typical
bandstructures which illustrates the crossover from the completely flat
Landau bands in the homogeneous magnetic field $\xi=\infty$, to the
bandstructure of electrons in a square lattice of Aharonov-Bohm
scatterers ($f=1/2$) at $\xi=0$. The bandstructures have been found by direct
numerical diagonalization of the Hamiltonian.
(See also Fig.~\ref{fig:onebands}).

In Fig.~\ref{sigmaXY} some typical results of the numerical calculations of
the Hall conductivity are shown.
In general we have no reason to expect, that the Hall
conductivity should be isotropic as a function of the angle between the
current and the flux vortex lattice. The results we present is for a current
running along the diagonal of the square lattice, i.e.\ along the $x$-axes
in Fig.~\ref{basis}. It is seen, that whenever there is a gap in the
spectrum, the Hall conductivity gives the quantized value in agreement with
the discussion in the last section. At Fermi energies not lying in a gap
$\sigma_{xy}$ allways tend to be lower than the value it has in the
homogeneous field. And in the limit $\xi \longmapsto 0$, $\sigma_{xy}$
vanishes alltogether. In this limit the electron sees a periodic array of
Aharonov-Bohm scatterers, each carrying half a flux quantum, and there is no
preferential scattering to either side. We observe that when the flat Landau
bands starts to get dispersion, the contribution to the Hall effect is no
longer distributed equally in the Brillouin zone. Instead it piles up across
local and global gaps in the spectrum resulting in the spiky $\sigma_{xy}$
spectra. Finally, we note that the amplitude of the fluctuations of
$\sigma_{xy}$ is of order $\frac{e^2}{h}$, consistent with this picture.

In the calculations presented in Fig.~\ref{sigmaXY}, the filling fraction is
limited to values below 30. This limitation comes from the fact, that we are
only able to handle matrices of limited size in the numerical calculations.
Filling fractions below 30 correspond to very low density electron gases, and
our calculations belong therefore to the ``quantum'' regime, i.e.\ to the
regime where $\lambda_F\gg \lambda_s$. According to the discussion in
Sec.~\ref{sec:single}, we expect a cross-over to a semiclassical regime,
for electron gases with higher density,  where
$\lambda_F\ll \lambda_s$, with a qualitatively different behaviour.

It is an important question wether it is possible to observe
these features of the Hall conductivity in experiments.
The conditions, which are nescessary, are that the mean free path
$l$ is long compared to all other lengths, and that $a,\lambda_s<\lambda_F$.
The last condition indicates that we are concerned with quantum magneto
transport,
in the sense that the vector potential is included in the proper quantum
treatment of the electrons. This is in contrast to the case
$\lambda_F \ll a,\lambda_s$ where the electron transport can be treated as the
semiclassical motion of localized wavepackets in a slowly varying magnetic
field.
Let us assume, in order to make some estimates, that the superconductor
has a London length somewhat less than 1000\AA, resulting in an exponential
length of the vortices $\lambda_s$ of about 1000\AA\- at the 2DEG, after the
broadening due to the distance between the superconducter and the 2DEG has
been taken into account. In order to have a variation in the
magnetic field we should have $a>\lambda_s$, and to be in the quantum regime
$\lambda_F>a,\lambda_s$.
This gives the estimate for the electron density
$n<6\cdot 10^{10}{\rm cm}^{-2}$, which
is not unrealistic. The effect of the impurities is (to first order), to give
the electrons a finite lifetime. This gives a finite longitudinal conductivity
$\sigma_{xx}$ and broadens the density of states. It also introduces
localized states at the band edges (Lifshitz tails). If the field is nearly
homogeneous, we have the standard quantum Hall picture with mobility edges
above and below every Landau band, resulting in the formation of plateaus in
$\sigma_{xy}$, which only can be observed at much higher magnetic
fields, of order $10^5$G, where the filling fraction is of order one.
On the other hand when the amount of impurities is low, that
is $k_F l \gg 1$, all these effects will be small, and we expect that
the features of $\sigma_{xy}$, shown in Fig.~\ref{sigmaXY}
will have observable consequenses in
$\rho_{xy}=\frac{\sigma_{xy}}{\sigma_{xx}^2 + \sigma_{xy}^2}$.

In Fig.~\ref{fig:onebands} we have plotted the bandstructure of electrons in
a square lattice of vortices carrying one flux quantum each. The calculation
has been made with the same basis as the bandstructures shown in
Fig.~\ref{fig:halfbands}, the only difference is that all fluxes have been
multiplied by a factor of 2, as can be seen from the spacing between the
Landau bands. In the limit where $\xi\rightarrow 0$,
and the vortex lattice becomes a regular array of Aharonov-Bohm scatterers,
we recover the well-known bandstructure of free electrons. This is in
agreement with our discussion of the AB-vortex in Sec.~\ref{sec:single}.

\subsubsection{Exchange of topological quanta}

An example of exchange of topological quanta between neighbouring bands is
shown in Fig.~\ref{exchange}.
The figure is an enlargement of the band structure around the $X$ point,
showing an accidential degeneracy between the 3'ed and 4'th band,
which occur about $\xi =0.035$. Also
indicated is the Hall conductance of each band in units of $\frac{e^2}{h}$,
found by numerical integration. At the degeneracy it is not possible to
define the Hall conductance for the individual bands. On the Brillouin zone
torus there are two $X$
points $X_1$, $X_2$, and the bands have a simple (1.\ order) degeneracy in
each. The numerical integration shows that two topological quanta are
transfered from the lower to the upper band,
and this is in full agreement with the discussion of Sec.~\ref{sec:bandcross}.
Exchange of topological quanta between bands is a common phenomena as $\xi$
is varyed, and this particular example has only been chosen as an illustration
of the general phenomena.

\subsubsection{Transverse conductivity in a disordered vortex phase}

When the vortices come from a thin film of superconducting material,
which have many impurities and crystal lattice defects acting as pinning
centers for the vortices,
the distribution of vortices will be disordered rather than forming a regular
Abrikosov lattice. The effect of the disorder will be
to wash out the destinctive features of the band structure, i.e.\ to
average out the characteristic fluctuations in $\sigma_{xy}$, leaving a smooth
curve in Fig.~\ref{sigmaXY} with a characteristic dimensionless
propertionality constant $s(\xi)$,
in the form $\sigma_{xy}=\frac{e^2}{h}s(\xi)\nu$, where $\nu$
is the number of electrons in the magnetic unit cell
$\nu=n a^2=n\frac{\phi_0}{B}$ (the filling fraction). With this conjecture
we can estimate the normalized Hall conductivity $s(\xi)$ by making a linear
fit to the calculated $\sigma_{xy}(\nu)$ distribution. In the experimental
situation $\lambda_s$ is constant, and this makes $s(\xi)$ a function
of the applied magnetic field through the relationship $\xi=\lambda_s/a
=\lambda_s\sqrt{B/\phi_0}$. In Fig.~\ref{fig:S} we have plotted $s(B)$
for a vortex exponential length $\lambda_s=80$nm.
The $s(B)$-curve shows essentially that the Hall effect of a
dilute distribution of vortices is strongly suppressed compared to the Hall
effect of a homogeneous magnetic field with the same average strength. This is
in good qualitative agreement with what is seen in the experiments of Geim et
al.~\cite{Geim92}.
When doing experiments, one is not directly measuring the conductivities,
but rather the resistivities $\rho_{xx}$, $\rho_{xy}$.
The experiments of Geim et al.\ cover the parameter range from
$\lambda_F\ll\lambda_s$ at high 2DEG densities, down to the  value
$\lambda_F /\lambda_s =0.7$ for the 2DEG with the lowest density
experimentially obtainable, where the new phenomena begin to occur.
Our numerical calculations belongs to the other side of this
cross-over where $\lambda_F \gg\lambda_s$.
The physical picture of this cross-over can be stated as follows.
On the high density side the magnetic field varies slowly over the size
of an electron wavepacket for electrons at the Fermi energy, with the result
that the wavepacket more or less behaves as a classical particle. On the
other side of the crossover $\lambda_F\gg\lambda_s$ the magnetic field varies
rapidly over the lengthscale of a wavepacket, for an electron at the Fermi
energy, and this introduces new phenomena of an essential quantum character.

\subsection{Superlattice potential}\label{sec:sup}

The general picture we have outlined so far of energy bands having
dispersion, with the dispersion giving rise to a non trivial behaviour of the
Hall conductivity, is not limited to the inhomogeneous magnetic field. The
dispersion could have another origin for instance a superlattice potential.
To illustrate this a series of calculations have been made on a 2DEG in a
homogeneous magnetic field, and a scalar potential which we have taken as a
square lattice cosine potential. This system have commensurability
problems because of the two ``interfering'' length scales, given respectively
by the magnetic length $l_B=\sqrt{\frac{\hbar}{eB}}$, and the period of the
superlattice potential $a$.
To make things as simple as possible we have fixed the
period of the cosine potential $a$, and the magnetic field strength $B$,
and only varied the amplitude of the cosine potential. Furthermore the
flux density of the magnetic field is tuned so that the flux through one unit
cell of the cosine potential is exactly one flux quantum.
The cosine potential is
\begin{equation}
U(x,y)=V_0(\cos 2\pi x/a + \cos 2\pi y/a),
\end{equation}
and the magnetic field $B=\phi_0/a^2$. The dimensionless parameter
controlling the shape of the energy band structure is in this case
\begin{equation}
v=\displaystyle{\frac{V_0}{\hbar\omega_c}}.
\end{equation}
Examples of the $\sigma_H$-spectra are shown in Fig.~\ref{super}.
It is observed that although the spectra look different from the vortex
lattice spectra, they have the same spiky nature. The spikes have the same
interpretation as in the vortex lattice system. Local spikes are due to
local gaps in the spectra. That is when to bands are close to each other
for some $\bbox{k}$ vector in the Brillouin zone, the result is a pile up of
topological charge across the gap, and this gives a spike in the
$\sigma_H$-spectra when the Fermi energy is swept across the gap.
Global spikes, that is spikes which go all the way up to the diagonal line
indicating the Landau limit, are due to global gaps in the energy spectrum,
combined with the topological quantization.

\section{Conclusion}

In Sec.~\ref{sec:single} of this paper we have considered the longitudinal
and the transverese resistivities of a 2DEG in a disordered distribution of
flux vortices, within the theoretical framework where each scattering event
is treated independently, and the electrons are non-interacting.
The general features observed in experiments are in agreement with the
results we have outlined, but we do not have quantitative agreement. If we
use the radius $R_v$ of the vortices as a fitting parameter, then the
longitudinal resistance fits to a radius of the order $R_v\simeq1000$nm, while
the transverse resistivity fits best at $R_v\simeq 30$nm.
The radius of the real vortices, is estimated by Geim to be $R_v\sim 100$nm.
This point requires analysis on a more elaborate level, in order to be
resolved.

In Sec.~\ref{sec:array} we have considered a new kind of experiment where a
2DEG is placed in a periodic magnetic field varying on a length scale
$\lambda_s$, comparable to (or less than) the Fermi wavelength $\lambda_F$
of the electrons. In this limit, where
it is nescessary to include explicitly the vector potential in a quantum
treatment of the electron motion, we expect the 2DEG to exhibit new
phenomena.
We have presented numerical results for a non-interacting 2DEG without
impurities showing characteristic spikes of the Hall conductivity versus
filling fraction, which can be understood in terms of local and global
energy gaps in the spectrum.

\section*{Acknowledgements}
Finally, we would like to thank Dr.\ Hans L. Skriver for computer time,
B. I. Halperin for drawing our attention to Ref. \cite{Axel}, to
A. K. Geim for showing us the experimental data prior to publication,
and to K. J. Eriksen, D. H. Lee, P. E. Lindelof, A. Smith and R. Taboryski
for stimulating discussions.

\begin{figure}
\caption{The scattering geometry for classical scattering on an idealized
cylindrical vortex with constant magnetic field inside and no field outside.}
\label{fig:geo}
\end{figure}

\begin{figure}
\caption{Differential cross section and classical trajectories for four
different values of the parameter $\gamma=l_c/R_v=$ 0.025, 0.40, 1.00, 2.50}
\label{fig:ccs}
\end{figure}

\begin{figure}
\caption{The geometry of the scattering situation.}
\label{fig:qgeo}
\end{figure}

\begin{figure}
\caption{The dimensionless Aharonov-Bohm cross section,
defined as ${\cal F}^{AB}(\theta)=\sum_l {\cal F}[\delta^{AB}_l]e^{il\theta}$.
Here plotted for $f=1/2$.
The different curves are cross sections corresponding to different values
of $\kappa$, and they have been translated relative to each other in order
not to overlap to much. The horisontal lines indicate the zero level for
the different curves. The lowest curve corresponds to $\kappa=0.25$, and the
other curves to respectively $1.00$, $2.50$, $5.00$, with $\kappa$
increacing upwards. The scale of the plot can be read off the distance
between the horisontal lines, which is equal to 1.}
\label{fig:AB}
\end{figure}

\begin{figure}
\caption{Differential cross sections plotted relative to the Aharonov-Bohm
cross section, for an electron scattering on a
magnetic flux vortex with finite radius. The curves have been obtained by
subtracting the dimensionless AB cross section from the calculated
finite radius cross sections. This have been done in order to single out the
effect of the finite radius. All plots are in the same scale, and this
includes the preceding figure, see caption of Fig.~4.
Furthermore all plots are with the same values
of $\kappa=$ 0.25, 1.00, 2.50, 5.00 increasing upwards.}
\label{fig:asymcross}
\end{figure}

\begin{figure}
\caption{Resistance efficiency of single vortex. Curves show $\zeta(\kappa)$
for different values of the flux fractions  $f$.}
\label{fig:zeta}
\end{figure}

\begin{figure}
\caption{Resistance efficiency $\zeta(\gamma)$, and Hall efficiency
factor $\alpha(\gamma)$ calculated from the classical
cross section for scattering on a magnetic flux tube. The parameter
$\gamma$ is given by the cyclotron radius divided by the radius
of the flux tube $\gamma=l_c/R_v$.}
\end{figure}

\begin{figure}
\caption{The efficiency of a dilute distribution of vortices in producing
Hall effect, compared to a homogeneous magnetic field with the same average
flux density. Curves show $\alpha$ as a function of $\kappa$ for different
values of the flux $f$.}
\label{fig:alpha}
\end{figure}

\begin{figure}
\caption{These plots of $\alpha$ and $\zeta$ for a vortex with $f=10$, shows
a striking structure of resonances at the values of $\kappa$ corresponding
to the Landau quantization energies.}
\label{fig:reso}
\end{figure}

\begin{figure}
\caption{(A) The unit cell with basis. The large circles indicates
the position of the Abrikosov vortices,
and the small circle indicate the position of the
Dirac vortex with the counter flux. The micro lattice shown here is
$6\times 6$, whereas all the numerical results we have presented are
obtained with a micro lattice of $10\times 10$ sites.
(B) Four concatenated unit cells,
showing the squarre lattice of Abrikosov vortices.}
\label{basis}
\end{figure}

\begin{figure}
\caption{The Brillouin zone with the conventionel symmetry labels.}
\label{Brillouin}
\end{figure}

\begin{figure}
\caption{Schematic energy band crossing, controlled by an outer parameter
$\gamma=\xi - \xi_0$.}
\label{bandcross}
\end{figure}

\begin{figure}
\caption{Bandstructures for 2D electrons in a square lattice of Abrikosov
vortices with $f=1/2$. The different plots show bandstructures
corresponding to various
values of the parameter $\xi$ equal to the ratio between the exponential
length of the magnetic field from a single vortex, and the lattice parameter.
The flux through a single vortex is half a flux quantum.}
\label{fig:halfbands}
\end{figure}

\begin{figure}
\caption{Calculated Hall conductivity versus filling fraction, for various
values of the ratio $\xi =\lambda_s /a$. These calculations are made on the
same system as the bandstructures of Fig.~3.4, that is a square lattice of
Abrikosov vortices with $f=1/2$. Each of the spectra are made as
follows. For 2000 equidistant Fermi energies $\epsilon_F$, the total Hall
conductivity $\sigma_H(\epsilon_F)$, and the integrated density of states
$\nu(\epsilon_F)$ are calculated numerically. This is done by 20 pages of
C++ code, running on a workstation for 24 hours. The $x$-axes indicates the
integrated density of states in units of filled bands. The $y$-axes indicates
the total Hall conductivity in units of the conductivity quantum $e^2/h$.
The diagonal line in the plots indicate the Hall conductivity in a
homogeneous magnetic field.}
\label{sigmaXY}
\end{figure}

\begin{figure}
\caption{The density of Hall effect, or ``topological charge'', plotted as
function of the filling fraction.
It is seen that for $\xi\rightarrow 0$ the distribution gets strongly
polarized, with a negative contribution to the Hall effect at the bottom part
of the bands, and a positive contribution at the uppermost part of the bands.}
\label{densit}
\end{figure}

\begin{figure}
\caption{Bandstructures for 2D electrons in a square lattice of vortices
carrying one flux quantum each, $f=1$.
The different plots shows bandstructures corresponding to various
values of the parameter $\xi$ equal to the ratio between the exponential
length of the magnetic field from a single vortex, and the lattice parameter.
The bandstructures have been calculated using the basis shown in Fig.~10,
with the only difference that
here the flux through each of the vortices are $\phi_0$, and the counter
Dirac flux is $-2\phi_0$.}
\label{fig:onebands}
\end{figure}

\begin{figure}
\caption{Exchange of topological quanta. The figure shows an enlargement of
the $f=1/2$ band structure around the $X$ point, where
a degeneracy between the 3'rd and 4'th band occur. (See Fig.~13).
The parameter $\gamma$ appearing in the figure
is defined as $\gamma=\xi - \xi_{0}$, with $\xi_{0}=0.035$.
The numbers give the Hall conductance of the  bands in units of
$\frac{e^2}{h}$, found by numerical integration.}
\label{exchange}
\end{figure}

\begin{figure}
\caption{The normalized Hall conductivity $s(B)$ for vortices of exponential
length $\lambda_s=$80nm.}
\label{fig:S}
\end{figure}

\begin{figure}
\caption{Calculated Hall conductivity versus filling fraction, for a 2DEG in
a homogeneous magnetic field, and a square lattice cosine potential, in the
special case where the magnetic flux density is exactly equal to one flux
quantum per unit cell area. The dimensionless parameter $v$ indicated in the
plots, is equal to the amplitude of the cosine potential divided by the
Landau energy $\hbar\omega_c$.
 The $x$-axes indicates the
integrated density of states in units of filled bands. The $y$-axes indicates
the total Hall conductivity in units of the conductivity quantum $e^2/h$.
The diagonal lines in the plots indicate the Hall conductivity in a
homogeneous magnetic field, without any potential.
For further details see the caption of Fig.~14}
\label{super}
\end{figure}

\end{document}